\documentclass[fleqn,usenatbib]{mnras}

\usepackage{newtxtext,newtxmath}
\usepackage{color}
\usepackage{multicol}
\usepackage{float}
\usepackage{lscape}

\usepackage[T1]{fontenc}

\DeclareRobustCommand{\VAN}[3]{#2}
\let\VANthebibliography\thebibliography
\def\thebibliography{\DeclareRobustCommand{\VAN}[3]{##3}\VANthebibliography}

\usepackage{graphicx}	
\usepackage{amsmath}	
\usepackage{ulem}
\usepackage{xcolor}

\newcommand{\naitl}{NaI(Tl)}

\newcommand{\keV}{keV\,}

\newcommand{\keVee}{\mbox{keV$_{\rm{ee}}$}\,}

\newcommand{\kevee}{keV$_{\rm{ee}}$\,}

\newcommand{\cms}{cm$^2$\,}

\newcommand{\gevc}{GeVc$^{-2}$\,}

\newcommand{\gevccm}{GeVc$^{-2}$cm$^{-3}$}

\newcommand{\counts}{[kg~keV~day]$^{-1}$\,}

\newcommand{\Msun}{$\,{\rm M}_{\odot}$}
\newcommand{\mf}{\textit{m12f}}
\newcommand{\kms}{km~s$^{-1}$}
\newcommand{\dru}{dru}

\newcommand{\Mpc}{Mpc}
\newcommand{\kpc}{kpc}
\newcommand{\pc}{pc}
\newcommand{\amu}{amu}
\newcommand{\supp}{\href{https://github.com/Grace-Lawrence/Gusts_in_the_Headwind-Uncertainties_in_Direct_Dark_Matter_Detection-Supplementary}{Supplementary}}
\newcommand{\DarkMaRK}{\href{https://github.com/Grace-Lawrence/Dark-MaRK}{Dark MaRK}}

\title[Uncertainties in Direct Dark Matter Detection]{Gusts in the Headwind: Uncertainties in Direct Dark Matter Detection}

\author[G. E. Lawrence et al.]{
Grace E. Lawrence,$^{1,2,3}$\thanks{E-mail: glawrence@swin.edu.au}
Alan R. Duffy,$^{1,2,3}$
Chris A. Blake,$^{1}$
Philip F. Hopkins$^{4}$
\\
$^{1}$Centre for Astrophysics and Supercomputing, Swinburne University of Technology, PO Box 218, Hawthorn VIC 3122, Australia\\
$^{2}$ARC Centre of Excellence for All Sky Astrophysics in 3 Dimensions (ASTRO 3D)\\
$^{3}$ARC Centre of Excellence for Dark Matter Particle Physics (CDM)
\\
$^{4}$TAPIR, Mailcode 350-17, California Institute of Technology, Pasadena, CA 91125, USA\\
}

\pubyear{2022}

\begin{document}
\label{firstpage}
\pagerange{\pageref{firstpage}--\pageref{lastpage}}
\maketitle

\begin{abstract}
We use high-resolution, hydrodynamic, galaxy simulations from the Latte suite of FIRE-2 simulations to investigate the inherent variation of dark matter in sub-sampled regions around the Solar Circle of a Milky Way-type analogue galaxy and its impact on direct dark matter detection. 
These simulations show that the baryonic backreaction, as well as the assembly history of substructures, has lasting impacts on the dark matter’s spatial and velocity distributions. These are experienced as `gusts' of dark matter wind around the Solar Circle, potentially complicating interpretations of direct detection experiments on Earth. 
We find that the velocity distribution function in the galactocentric frame shows strong deviations from the Maxwell Boltzmann form typically assumed in the fiducial Standard Halo Model, indicating the presence of high-velocity substructures. 
By introducing a new numerical integration technique which removes any dependencies on the Standard Halo Model, we generate event-rate predictions for both single-element Germanium and compound Sodium Iodide detectors, and explore how the variability of dark matter around the Solar Circle influences annual modulation signal predictions. 
We find that these velocity substructures contribute additional astrophysical uncertainty to the interpretation of event rates, although their impact on summary statistics such as the peak day of annual modulation is generally low.
\end{abstract}
\section{Introduction}
The nature of the dominant, non-luminous, weakly interacting component of our universe, dark matter, remains one of the primary unanswered questions in modern astrophysics. 
Evidence for dark matter's existence is widespread in observational astronomy, suggested first by \cite{1933AcHPh...6..110Z,1937ApJ....86..217Z}, who found that the velocity dispersion of galaxies within the Coma cluster was too high for the cluster to remain bound given observational mass measurements \citep{Bertone_2010}, and then by rotational velocity curves demonstrating the unexpectedly fast rate of rotation in the outer regions of spiral galaxies, indicating a larger portion of mass contained in these regions, e.g. \cite{1970ApJ...159..379R}.

Dark matter's existence has been further confirmed across many scales, from dwarf galaxies \citep{Ackermann_2015} to galaxy clusters \citep[e.g.][]{Markevitch_2004}, to the large scale structure of the universe~\citep{10.1111/j.1365-2966.2006.11269.x,Tojeiro_2014}. 
Fluctuations in the cosmic microwave background, the baryon acoustic oscillations, reveal the percentage makeup of the baryonic and dark constituents of our universe, with baryons comprising 4.9\%  and dark matter 26.4\% \citep{Planck_2018}.
Presently, dark matter can only be studied via its gravitational effects on observable objects.
This indirect observational evidence is also supported by cosmological simulations which require a dark matter term to accurately reproduce the observed universe \citep*{Navarro_1997,2005Natur.435..629S,10.1111/j.1365-2966.2009.16029.x}.

Globally, there are many efforts underway, both direct and indirect, to detect and characterise the nature of dark matter. 
Direct detection efforts aim to identify the signature left when a dark matter particle of mass $M_{D}$ and a target detector nuclei, with a reduced target mass of $M_T$, undergo an elastic collision and exchange energy. 
The target atom releases this energy in a potentially observable manner, with the solar system's circumnavigation of the galaxy providing a near-constant expected dark matter flux from the direction of the Cygnus constellation. 
Furthermore, such searches may then observe a secondary signal through the differential motion of the Earth itself around the Sun, giving rise to an annual modulation in the flux of dark matter particles through the Earth - referred to as the dark matter headwind \citep{PhysRevD.33.3495}.

An isotropic velocity distribution of dark matter particles will give rise to a sinusoidal signal peaking in June \citep{Green:2003yh}, which will be visible above the background radiation contaminants (such as K-40 and cosmogenically activated Na-22, which will not exhibit a strong seasonal dependence; \citealt{bolognino2020direct}).
If the source of the signal is truly astrophysical, and by extension considered to be dark matter, then this fluctuation should maintain phase regardless of the hemisphere the experiment is conducted in.

Numerous experiments are currently operating globally, and coming online in the near future, with the sensitivity to probe physically meaningful regions of parameter space \citep{Froborg_2020}. 
Notably, one experimental effort, the DAMA/LIBRA collaboration, has a longstanding claim of detecting the annual modulation of dark matter \citep*{2008EPJC...56..333B, Bernabei_2018,2021arXiv211004734B}. However this is at tension with other global experiments which report null signals in the parameter space spanned by the DAMA claim.

Experiments to test the DAMA claim are already underway. The Annual modulation with \naitl \ Scintillators (ANAIS) experiment, using nine Sodium Iodide (NaI) crystals \citep{PhysRevLett.123.031301,Amar__2020} finds results consistent with the null hypothesis of no modulation \citep{Froborg_2020}. 
The Collaboration Of Sodium IodiNe Experiments (COSINE, \citealt{PhysRevLett.123.031302}), another crystal Sodium Iodide experiment similar in design to DAMA, reports an early result which is consistent with both the null hypothesis and DAMA's 2-6\kevee best fit. 

The SABRE experiment (Sodium Iodide with Active Background Rejection; \citealt{Bignell_2020}) offers a decisive opportunity to test the astrophysical nature \citep{Froborg_2020} of any annual modulation with its dual-hemisphere design, that sees similar (but much higher purity) thallium-doped Sodium Iodide crystals to DAMA, deployed in Italy and Australia.

The amplitude and phase of this sinusoidal signal are key criteria in successfully identifying a dark matter signal from Earth. 
Generally accepted criteria for identifying an annual modulation signal due to dark matter are;
\begin{enumerate}
    \item The phase of the annual modulation signal should peak in the middle of the year, regardless of which hemisphere the experiment is operating in, indicating the signal is a result of the Earth's motion through the dark matter `headwind'. In this way we can distinguish between a local, seasonal, modulation result, and an astrophysical signal. (The phase can invert at low recoil energies, as will be discussed later).
    \item The amplitude of the annual modulation signal should not vary by more than 10\% (the variance of the Earth's velocity around the Sun). This amplitude can be more precisely estimated with a thorough understanding of the input parameters and uncertainties of the dark matter model, dark matter velocity distribution in the halo and the experimental hardware.
    \item There should be a strong directional dependence from the dark matter headwind, offering a potential for further insights using directional dark matter detectors \citep{Mayet_2016}.
\end{enumerate}
This paper explores how the inherent variability in dark matter environments around the Solar Circle can give rise to non-negligible changes in direct detection parameters, like amplitude and phase.

The fiducial description of dark matter in our Milky Way, referred to as the Standard Halo Model (SHM), assumes it exists as a single-component, cored, isothermal sphere of dark matter particles \citep{ PhysRevD.99.023012}, parameterised by a central density and core radius following a Navarro$\mbox{-}$Frenk$\mbox{-}$White (NFW) density profile \citep{Navarro_1997}. 
The collisionless Boltzmann equation, which expresses the flow of particle points throughout phase-space, is solved using the $\ \rho \propto r^{-2}$ density profile (assuming an isotropic spread of velocities). 
This isothermal profile is a reasonable approximation for the NFW profile in the solar neighbourhood, and together result in the Maxwell Boltzmann equation \citep{quantum_slides} for individual particles' properties. 
This is traditionally taken as the assumed velocity distribution of dark matter within the Solar Circle, with a manual truncation at the Milky Way's galactic escape speed.

In the absence of observational constraints, the most reliable alternative to inform the accuracy of this assumption is using hydrodynamic simulations.
Significant departures from this SHM have been quantified in previous works \citep{Savage_2006,Kuhlen_2010, GREEN_2012,Kelso_2016,Necib_2019,Lacroix_2020} and this work will further investigate how these assumptions, along with different dark matter particle masses, can have a significant influence on the expected detection rates for WIMP (Weakly Interacting Massive Particle) dark matter, revealed through simulated Milky Way analogue halos.

Recently numerical simulations have been used to quantify the velocity distribution of dark matter in the Milky Way \citep{Bozorgnia_2017}. The effects of the simulations' anisotropic velocity space structure manifests as shifts in the peak day of $\approx$ 20 days for samples about the Solar Circle.  
\cite{Green:2003yh} found that overly simplistic assumptions about the Earth's motion about the Sun and through the Milky Way lead to errors of up to ten days in the phase of the expected signals and up to tens of percent in the shape of the signal, even when assuming an isotropic velocity distribution. With an observationally-motivated velocity distribution, this phase change increases to up to 20 days.    
Work by \cite{Pillepich_2014} using the Eris simulation \citep{Eris} found a contraction of baryons can pull the dark matter into the disk plane without forcing it to co-rotate and that accretion and disruption of satellites can result in a dark matter component with a net angular momentum. 
The concentration of dark matter in the centre of a galaxy from the weak dark disk acts to increase the density and subsequent time-averaged scattering rate by a few percent at low recoil energies. 
However, at high velocities, the baryonic contraction creates a strong enhancement in the scattering rates.
However, \citet{Schaller_2016} uses the APOSTLE project \citep{10.1093/mnras/stw145,10.1093/mnras/stv2970} to find that the presence of these dark disks are rare.
Additionally, observational data from the Sloan Digital Sky Survey and latest Gaia release have also been used to trace the dark matter distribution by using Metal-Poor stars as a proxy \citep{2018PhRvL.120d1102H, Necib_2019}. This work found a lower peak speed and smaller dispersion in the velocity distribution when compared to the Standard Halo Model. The results also found the distribution to not be isotropic as assumed in the SHM.

\cite{O_Hare_2020} use of Gaia satellite data to identify `Dark Shards' containing substantial stellar streams resulted in modifications of fundamental properties for expected dark matter signals. The consequential departures of the speed distribution of dark matter in the solar neighbourhood from SHM assumptions caused shifts in the peak day of predicted annual modulation signals caused by nuclear recoils.

This work aims to provide insight into how realistic velocity distributions from the highest resolution TreePM+MLFM* \footnote{PM: particle-mesh; TreePM: tree + PM; MLFM: mesh-free finite mass \citep{2020NatRP...2...42V}} hydrodynamic zoom-in simulations will influence the predicted direct detection signals, and the interpretation of the measured signals. 
In particular, we evaluate the error budget for analysis, most notably for the phase ($t_0$) and amplitude ($S_m$) parameters, with an emphasis on sample variance depending on location about the Solar Circle, and the type of dark matter environment that the Earth is passing through. 
Motivated by a more realistic exploration of the dark matter distribution close to the Sun's orbit and the potential time-dependent structures that may persist in these galaxy realizations, hydrodynamic simulations with detailed galaxy formation models are used to probe the internal structure of dark matter halos, and to create bespoke predictions for terrestrial dark matter detectors given the distribution of particles in the simulation. 
The use of high resolution hydrodynamic simulations allows these parameters and assumptions to be investigated and constrained in the absence of experimental data.
Calculations are computed using the \textit{Dark MaRK} package, presented in a follow up paper (Lawrence {\it in prep}) and available on Github \footnote{\url{https://github.com/Grace-Lawrence/Dark-MaRK}}.

This paper will focus on the WIMP candidate for dark matter, motivated by the WIMP miracle and extensions to the standard model \citep{Bertone_2010}. 
Exploring WIMP models between $1-100$\gevc, we seek to inform direct detection searches focused on this candidate, through the process of nuclear recoil. 
This work will focus on commonly used target detectors of materials based around Germanium and Sodium Iodide. In particular, the choice of the latter target material is due to its use by the DAMA collaboration, as they continue to be the only claimed detection of a dark matter signal, with 9.5$\sigma$ \citep{Bernabei_2018}.

The remainder of the paper is organised as follows. 
In Section \ref{sec:lattesims} we describe the Latte simulations used in the work and provide details of the chosen \mf\ halo, in addition to a comparison of galaxy properties to our own Milky Way. 
In Section \ref{sec:frame} we present the process of sampling our halo as well as the subsequent frame transformations to convert our data into relevant reference frames. 
In Section \ref{sec:rate_calcs} we detail the rate calculation equations used in this work, and in the \textit{Dark MaRK} package. 
In Section \ref{sec:Results} we present results for both Germanium and Sodium Iodide detectors and in Section \ref{sec:Discussion} we discuss the results and their impact on experimental interpretations. We conclude with Section \ref{sec:conclusions} and outline future work.

\section{Simulations}\label{sec:lattesims}
The Latte suite of FIRE-2 cosmological zoom-in baryonic simulations of Milky Way-mass galaxies \citep{https://doi.org/10.48550/arxiv.2202.06969,Wetzel_2016}, part of the Feedback In Realistic Environments (FIRE) simulation project, were run using the GIZMO gravity plus hydrodynamics code in meshless finite-mass (MFM) mode \citep{2015MNRAS.450...53H} and the FIRE-2 physics model \citep{10.1093/mnras/sty1690}. The Latte simulations model the formation of Milky Way-mass halos to the present day within the $\Lambda$CDM cosmology. These hydrodynamic simulations include dark matter, gas and star particles to model the stellar disk, stellar halo and dark matter halo of these systems \citep{10.1093/mnras/stu1738, Wetzel_2016,10.1093/mnras/stx1710,10.1093/mnras/sty1690}.

These simulations are run using the FIRE-2 model for star formation/feedback and GIZMO, the flexible, massively-parallel, multi-physics simulation code, descended from GADGET \citep*{2001NewA....6...79S,2015MNRAS.450...53H,10.1093/mnras/sty1690}.  
GIZMO uses a TREE+PM gravity solver and mesh-free finite-mass method for adaptive spatial resolution \citep{Wetzel_2016}. 
Halos in this suite are selected from a cosmological volume of periodic box length $85.5$\Mpc\ with $\Lambda$CDM cosmology given by $\Omega_{\Lambda}$=0.728, $\Omega_{\rm m}$=0.272, $\Omega_{\rm b}$=0.0455, $h$=0.702, $\sigma_8$=0.807. 
Halos are also selected using an isolation criteria such that they have no neighbouring halos of similar mass within $<5 R_{200}$, where $R_{200}$ is the virial radius at which the average density within is 200 times the critical density of the universe.
The particle mass resolution is 35000\Msun~for dark matter particles and 7070\Msun for gas and star particles. 
The spatial (force) resolution is 40 \pc\ for dark matter, 4 \pc\ for stars, and 1 \pc\ (minimum) for gas.

\begin{figure}\label{fig:halo_realization}
 \begin{center}
  \setkeys{Gin}{width = 0.70\linewidth}

  \includegraphics{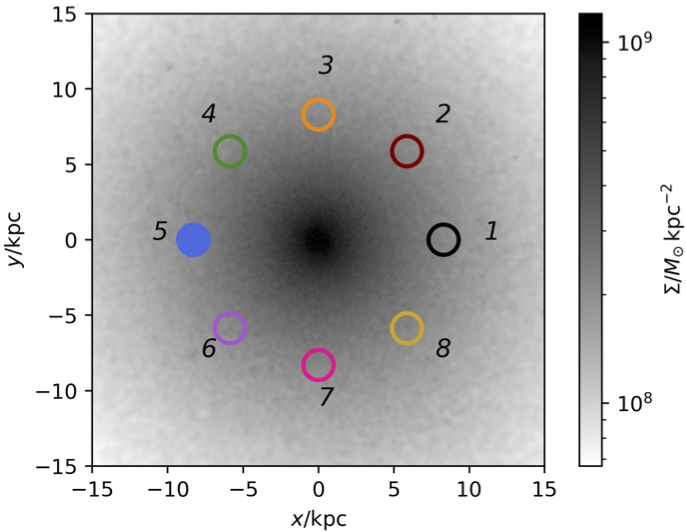}\,%
  \includegraphics{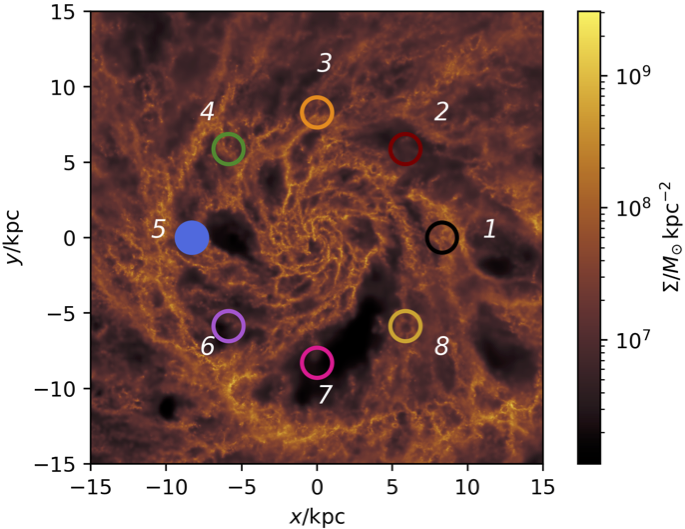}\,%
  \includegraphics[width=61mm]{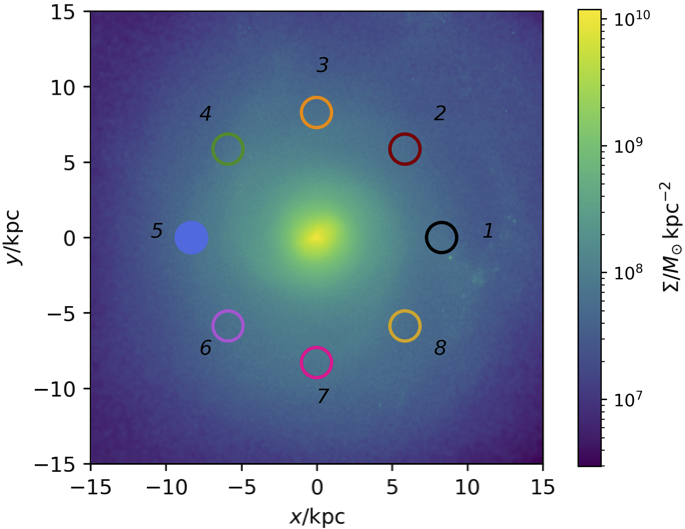}
  
  \includegraphics{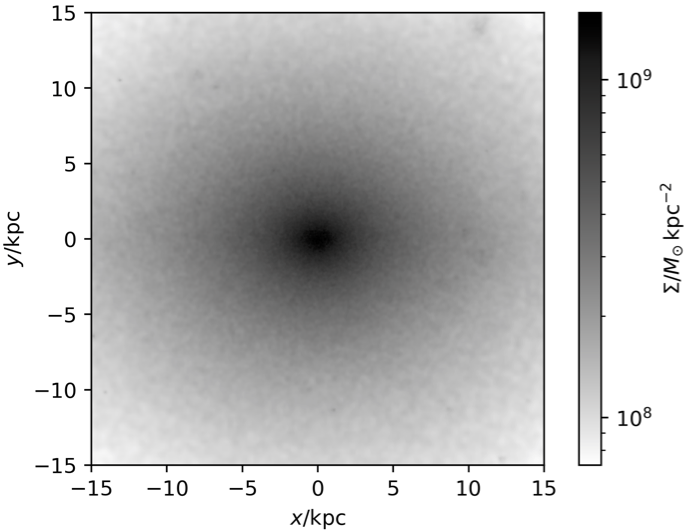}\,%
  \includegraphics{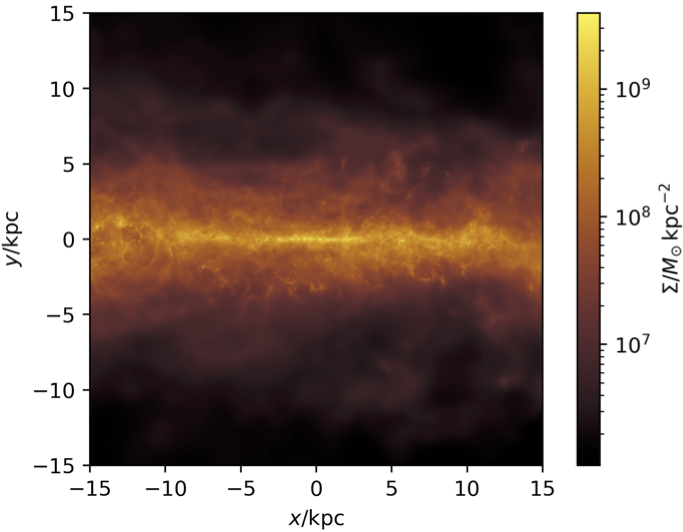}\,%
  \includegraphics[width=0.71\linewidth]{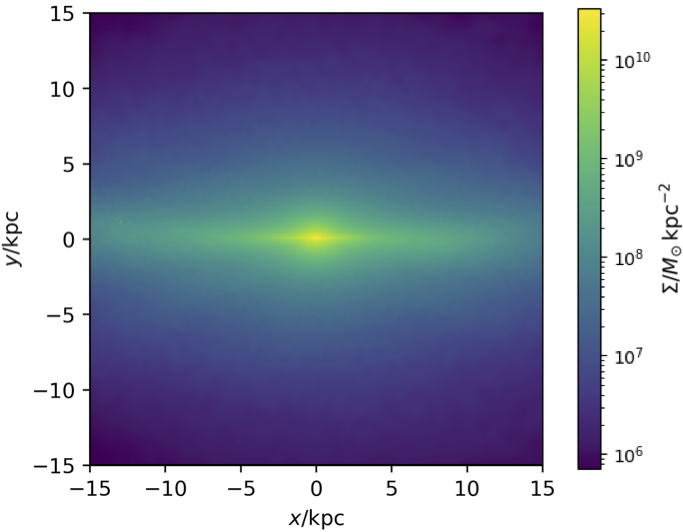}
\caption{Face-on (top) and side-on (bottom) profiles of the dark matter (left), gas (centre) and stars (right) components of the \mf\ halo from the Latte suite of simulations \citep{Wetzel_2016}. The colored circles denote the eight samples used in this work (to scale), each of radius 1\kpc. The filled circle represents the Earth's location in the Milky Way, to which each circle is rotated. The colours corresponding to each sample are consistent throughout this work.}
\end{center}
\end{figure}

From this suite, we select the Milky Way analogue, halo \mf\ (first introduced in \cite{10.1093/mnras/stx1710}). 
A spiral galaxy with an extended disk and a stellar stream within the solar neighbourhood makes \mf\ an excellent proxy environment with which to test the variability of dark matter around a realistic, Milky Way halo analogue. 
Resolved with approximately $96\times 10^6$ dark matter particles, $80\times 10^6$ gas particles and $16\times 10^6$ star particles (see Figure \ref{fig:halo_realization}), \mf\ was closest to the Milky Way in terms of stellar mass and size. It contains 8 stellar streams within the galaxy \citep{2021ApJ...920...10P}, with one contained in the solar neighbourhood, our region of interest \citep{Sanderson_2020}. Analogous to the many stellar streams identified in the Milky Way \citep{Malhan_2018} and the notable Gaia Enceladus \citep{Helmi_2018}, the inclusion of stellar streams and debris flows in the astrophysical considerations of direct detection is an important inclusion to ensure realistic expectations \citep{PhysRevD.99.023012,O_Hare_2018,Necib_2019,O_Hare_2020}. 
Though streams are present within the solar region, there is no overlap of stream particles, as identified in \cite{2021ApJ...920...10P}, and the sampled regions of interest in this work.

The simulated halo \mf\ has a virial mass of $\ 1.58\times 10^{12}$\Msun\ \citep{10.1093/mnras/sty2513} compared to the Milky Way's $0.96^{+0.29}_{-0.28}\times 10^{12}$\Msun\ \citep{Patel_2018}. 
The rotational velocity for \mf\ peaks at $\sim 270$ \kms, and at the solar radius (defined at 8.3\kpc), the circular velocity of$\ \sim250$ \kms\ is in close alignment with the Milky Way's circular velocity of $\sim 230$ \kms \citep{Eilers_2019}. However, this small velocity difference will be corrected for as discussed in Section~\ref{sec:frame}.
The halo has a marginally disturbed gas disk at the present day \citep{10.1093/mnras/sty2513}, resulting from a recent interaction with a gas-rich sub-halo, and a large tidal stream between $15-25$ \kpc\ \citep{Sanderson_2020}. This type of coherent velocity substructure makes the galaxy simulation a particularly useful analogue to test the potential impacts of such structures on direct detection experiments by selecting regions which contain such structures in the sub-samples (as discussed below). 
In summary, the simulated halo \mf\ is similar in terms of stellar and gas mass, size and stellar morphology to the Milky Way, and we refer readers to \citet{Sanderson_2020} for a more detailed comparison.

\section{Frame Transformations}\label{sec:frame}
\begin{table}
\centering
\begin{tabular}{|c|c|c|c|c|c|}
\# & $p_x$ & $p_y$  & $\theta_z (^{\circ})$ & $N_P$ & $\rho_{DM}$ \\
\hline
1 & 8.300 & 0.000   & 180 & 1392 & 0.44  \\
2 & 5.869 & 5.869   & 225 & 1452 & 0.46 \\
3 & 0.000 & 8.300   & 270 & 1459 & 0.46 \\
4 & -5.869 & 5.869  & 315 & 1409 & 0.45 \\
5 & -8.300 & 0.000  & 0 & 7233 & 2.29 \\
6 & -5.869 & -5.869 & -45 & 1460 & 0.46 \\
7 & -0.000 & -8.300 & -90 & 1448 & 0.46 \\
8 & 5.869 & -5.869  & -135 & 1399 & 0.44  \\

\hline
\end{tabular}
\caption{The centre coordinates of each of the 8 samples (in \kpc),  with a radius of 1\kpc\ and $p_z = 0.027$ \kpc. All samples are rotated back to the Earth's position, Sample 5, using rotation angle $\theta_z$ where the origin at (0,0,0)\,\kpc\, is assumed to be the galactic centre (GC). $N_P$ represents the number of simulation particles in each sample. The dark matter particle mass resolution of 35000\Msun\ \citep{Sanderson_2020} is used to calculate the densities, $\rho_{DM}$ in \gevccm.}\label{table:samp_coords}
\end{table}

We now describe the method by which the sub-sampled regions from the simulation are chosen, and transferred into event rates in the lab frame for idealised direct detection experiments. We use the python package Pynbody~\citep{pynbody} to translate the object into the galactocentric reference frame. The Pynbody routine, using the angular momentum vector, centres the simulation and rotates its axis such that the disk lies in the x-y plane and the centre is at the coordinate origin. This is then inverted in the x-coordinate plane to create a left-hand centric system\footnote{The system, with its origin at the galactic centre, has the x-axis positive in the direction of the galactic centre to the Sun. The y-axis is positive in the direction of galactic rotation with the z-axis positive perpendicular to the galactic plane.}. 

We then sub-sample eight, evenly-spaced spherical samples of radius 1\kpc\ about the Solar Circle, a selection volume sufficiently large to ensure a representative sample of particles. 
As well as the individual samples, we also create a Solar Circle sample by stacking all 8 sub-samples to create a sample of 17252 particles, which have all been rotated back to the solar system's co-ordinates.

Table \ref{table:samp_coords} summarises the statistics of each sample. We state the coordinates of these samples in the galactocentric reference frame, ${f}_{gal}$, and then rotate them to the Earth's position at $(-8.3,\ 0,\ 0.027)\,$\kpc\ (equivalent to the centre coordinates of Sample 5) by an amount $\theta_z$, given in Table \ref{table:samp_coords}, to simplify comparison of the sample properties. Column 6 in Table \ref{table:samp_coords} shows the densities calculated for each sample, where the dark matter simulation particles have a mass $35000$\Msun. These values are listed to demonstrate the variation among the samples. Sample 5 has a significantly higher density than the others; not associated with a feature of the inherent variation of dark matter around the Solar Circle (see Figure \ref{fig:Rho_All_Samples} in Appendix \ref{Appendix:VDF} for density histograms for each sample).\newline
\newline
\newline
\newline
\newline
\newline
\newline
\newline 
\newline
\newline
\newline
\newline
\newline
\newline
\newline
\newline
\newline
\newline
\newline
\newline
\newline
\newline
\newline
\newline
\newline
\newline

We caution the reader that the galaxy properties of \mf, while analogous to the Milky Way, are not identical. 

In order to account for the difference between the observationally measured $\vec{v}_{Earth}$ and the speed of the simulation particles given by the circular velocity of the halo, we perform a modest correction.
The simulation velocities are boosted in the galactocentric frame, $f_{gal}$, to align with the observational velocities of the Earth through the solar system, as specified by the \cite{astropy:2013,astropy:2018}. 
In practice, a general boost of $(-1.27, -23.29, +2.31)$ \kms is imparted to all particle velocity vectors after their sample has been rotated to the solar system's location.
This boost of the velocity vector has only a minor impact on event rates with $\ f_{geo}(\vec{v}, t)$, but ensures that these results are more directly calibrated to terrestrial dark matter detection experiments.

The distribution of velocity vectors in the galactocentric reference frame, ${f}_{gal}(\vec{v})$, are then transformed into the distribution in the geocentric (lab) reference frame, ${f}_{geo}(\vec{v}, t)$, via a Galilean boost \citep{McCabe_2014}
\begin{equation}
    f_{geo}(\vec{v}, t) = {f}_{gal}(\vec{v}+\vec{v}_{Earth}(t)),
\end{equation}
where $\vec{v}$ is the simulation velocity and the velocity of the Earth, $v_E$, with respect to the galactocentric rest frame is
\begin{equation}
    \vec{v}_{E}(t) = \vec{v}_{LSR}+\vec{v}_{pec}+\vec{u}_E(t),
\end{equation} 
using the Astropy coordinate transformation \citep{astropy:2013,astropy:2018}, where $\vec{v}_{LSR}$ is the local standard of rest, $\vec{v}_{pec}$ is the peculiar motion of the Sun with respect to the local standard of rest and $\vec{u}_E(t)$ is the Earth's velocity as it orbits the Sun. Conventionally, $v_{LSR} = (0,220,0)$ \kms and $v_{pec} = (11.1,12.2,7.3)$ \kms \citep{Sch_nrich_2010}.
This velocity distribution, $f_{geo}$, as seen from the detector on Earth, is used to calculate the nuclear recoil energy spectrum, $\frac{dR}{dE_R}$, as discussed in Section \ref{sec:rate_calcs}. 
This distribution will impact the expected event rate and annual modulation signals for terrestrial dark matter detectors.

\begin{figure}
    \centering
    \includegraphics[width = \linewidth]{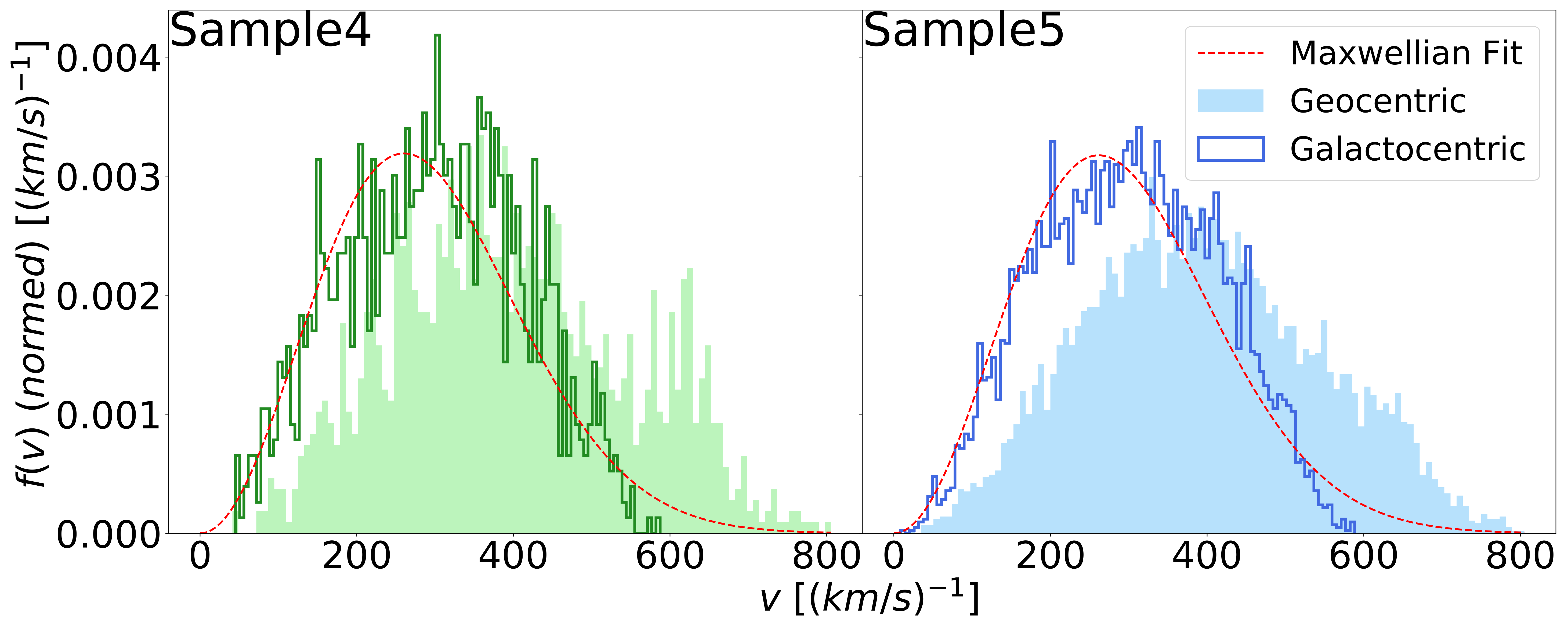}
    \includegraphics[width = \linewidth]{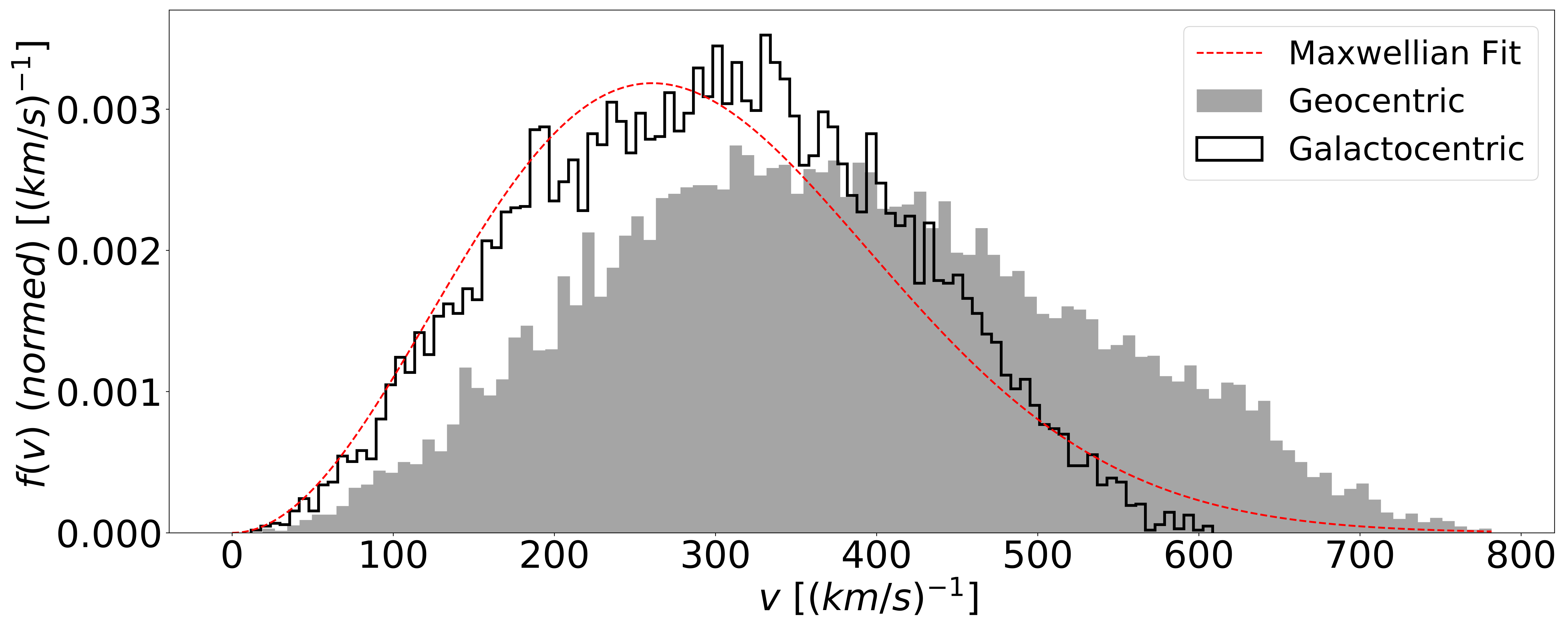}
    \caption{Top: Velocity distribution functions for Sample 4 and Sample 5 showing the galactocentric (open histogram, thick line) and geocentric (filled histogram) distributions, with a Maxwell Boltzmann function fit (dashed line) to the galactocentric distribution in the left and right images, respectively. Bottom: The total velocity distribution function for the Solar Circle (achieved by stacking the 8 individual solar samples).}
    \label{fig:VDF_Solar_Circle}
\end{figure}

In Figure \ref{fig:VDF_Solar_Circle} we show the velocity distribution function (VDF) for Sample 4 and Sample 5, selected to demonstrate the range of velocity distributions across the Solar Circle samples. 
Sample 4 clearly demonstrates significant substructure in velocity space, providing a vivid example of how inhomogeneous and `messy' a sample can be.
This includes significant substructure in the high-velocity tail, which would not be present in the commonly-assumed Maxwellian distribution used in the literature.
Sample 5, on the other hand, is smoother and better described by a Maxwellian distribution, more akin to the standard fiducial theoretical assumptions \citep{PhysRevD.33.3495}. This sample exhibits significantly higher particle density, in contrast with the remaining samples which match with theoretical expectations of the Milky Way halo density at the Solar Circle (see Table \ref{table:samp_coords}).

We note that while past works utilizing an analytical model for the VDF would impose a sharp cut-off to the distribution at the escape speed of the galaxy \citep{necib2022substructure}, the VDFs in this work are representative of the particles which, according to the simulation's halo, have remained within the galaxy. This means that the high velocity tails are present in the halo, and that the fast moving substructure is gravitationally bound to the halo at the time of the snapshot. This inherent structure, particularly the high velocity tail, is evident in 7 of the 8 samples, which can be viewed in Appendix \ref{Appendix:VDF}. This significant high-velocity structure highlights the extensive deviations of the dark matter structure in velocity space around the Solar Circle that is inherent to a `messy' galaxy, and not directly attributed to a stream or debris flow. Stacking the particles from all samples to create a proxy for the Solar Circle in Figure \ref{fig:VDF_Solar_Circle}, we see that while some of the high-velocity fluctuations are smoothed, a bulge is present in the high-velocity tail and there is significant deviation from the Maxwellian fit. These velocity distributions affect the fly-through and detection rates of direct dark matter detectors through Equation \ref{eq:rate_sum}, as explained in Section \ref{sec:rate_calcs}. 

This intrinsic variability in the flux of dark matter through the Earth as it orbits about the Sun circumnavigating the galactic centre is evident in the annual modulation signal, which can be parameterised by fitting a sinusoidal function \citep{2008EPJC...56..333B}
\begin{equation}\label{eq:DAMA_fit}
    S_i(E) = S_0(E)+S_m(E) \cos{[\omega (t_i-t_0)]}.
\end{equation}
Where $\omega$ is the angular frequency of a year $\frac{2\pi}{T(yr)}$ with period T.
We fit the parameters of this equation -- the overall rate $S_0$, the overall rate change $S_m$ and the phase / peak day $t_0$ -- to the event rate data using a non-linear least squares method \citep{2020SciPy-NMeth}.
Confidence intervals for these signals were evaluated using a bootstrap resampling technique, where the VDFs are randomly sampled (with replacement of each particle) 10,000 times.
Rate calculations were then performed for each individual realization, providing 10,000 annual modulations that we used to compute the confidence intervals.  High-resolution simulations give the advantage of a greater number of particles which provide tighter constraints on the fitted parameters. 

\section{Rate Calculations}\label{sec:rate_calcs}
The velocity distribution in the lab frame has direct consequences for the spectral function $\frac{dR}{dE_R}$ which describes the differential event rate of dark matter detection per unit recoil energy, $E_R$.  This velocity distribution function is integrated over a velocity range from $v_{min}$ to $v_{max}$ to generate the spectral function. Here, $v_{min}$ is the minimum detectable velocity for a dark matter particle of certain recoil energy given by $v_{min} = \sqrt{\frac{2E_r}{r M_{D}}}$ where $r$ is the kinematic factor for collisions given by $r = \frac{4M_DM_T}{(M_D+M_T)^2}$. 
In practice $v_{max}$ is the escape speed of the galaxy, which truncates the galactocentric speed distribution. But in our simulated galaxy only particles that are gravitationally bound are selected in the definition of the halo so this upper limit has no effect on those simulation results (but will when testing the Maxwell Boltzmann distribution which is formally unbounded otherwise).
This spectral recoil function is modulated by $v_E$ and the subsequent seasonal variation in these curves gives rise to a crossing point where the amplitude of the sinusoidal function will turnover. 

The differential event rate per unit detector mass, $dR$, for a beam of dark matter
particles with speed $v$, incident on a target of atomic mass $A$ with
interaction cross-section per nucleus $\sigma$, is given by
\begin{equation}\label{eq:dR}
  dR = \frac{N_0}{A} \, \sigma \, v \, dn \,,
\end{equation}
where $N_0$ is Avogadro's number and $dn$ is the differential dark matter particle density for particles with this particular speed. The densities for each sample are listed in Table \ref{table:samp_coords} however this work takes a fiducial value of 0.3\gevc cm$^3$. This $\rho_D$ value will scale the event rates without influencing the phase of the annual modulation signal. The cross section for spin-independent interactions at zero-momentum transfer is normalized to the WIMP-nucleon cross-section where 
\begin{equation}
\sigma = \frac{\mu^2}{\mu^2_p}A^2 \sigma_p \,,
\label{eq:sigma}
\end{equation}
where $\mu = \frac{M_DM_T}{M_D+M_T}$, $\mu_p$ is the WIMP-proton reduced mass and $\sigma_p$ is the scattering cross-section with a proton \citep{Kelso_2016}. We treat interactions as spin-independent (meaning the same for protons and neutrons). This interaction cross-section is altered by the Form Factor (Section \ref{sec:Form_factor}, Equation \ref{eqn:sigma}) to introduce a velocity dependence. For low momentum transfers these add in phase to give an interaction term of $A^2$ where A is the atomic mass of the detector material.

The differential particle density within a velocity element $d^3\vec{v}$
is given by
\begin{equation}
  dn = \frac{n_0}{k} \, f_{geo}(\vec{v},\vec{v}_E) \, d^3\vec{v}\,,
\end{equation}
where $f(\vec{v},\vec{v}_E)$ is the velocity distribution of particles
in the Earth's frame, which is a function of the particle velocity
$\vec{v}$ and Earth's velocity $\vec{v}_E$, $n_0$ is the local number
density of dark matter, and $k$ is a normalisation constant for the
velocity distribution, such that
\begin{equation}
  k = \int f_{geo}(\vec{v},\vec{v}_E) \, d^3\vec{v}.
\end{equation}
Assuming isotropic scattering in the centre of mass frame, the nuclear recoils in the target (for
dark matter particles of given speed $v$) are uniformly distributed in recoil
energy $E_R$ over the range $0 \le E_R \le E \, r$, where $E =
\frac{1}{2} M_D v^2$ is the incident kinetic energy of the dark matter
particle with mass $M_D$ and $r$ is the kinematic factor for the
collisions as before.  Hence, the event rate integrated over velocity, per unit
recoil energy, is given by
\begin{eqnarray}\label{eq:dRdEr_derive}
  \frac{dR}{dE_R} &=& \int_{v_{min}}^{v_{max}} \frac{dR}{E \, r}, \\ &=& \int_{v_{min}}^{v_{max}} \frac{N_0}{A}
  \, \sigma \, v \, \frac{1}{\frac{1}{2} M_D v^2 \, r} \,
  \frac{n_0}{k} \, f_{geo}(\vec{v},\vec{v}_E) \, d^3\vec{v}, \\ &=& \frac{2
    \, N_0 \, n_0 \, \sigma}{A \, k \, M_D \, r} \int_{v_{min}}^{v_{max}} \frac{1}{v} \,
  f_{geo}(\vec{v},\vec{v}_E) \, d^3\vec{v}.
\end{eqnarray} 
In order to evaluate this integral as a sum over $N$ simulation
particles labelled by $i$, we make the replacement $\frac{1}{k} \int
d^3\vec{v} \, f_{geo}(\vec{v},\vec{v}_E) \rightarrow \frac{1}{N} \sum_i$, and
hence
\begin{equation}\label{eq:rate_sum}
  \frac{dR}{dE_R} = \frac{2 \, N_0 \, n_0 \, \sigma}{A \, M_D \, r}
  \frac{1}{N} \sum_{v_{min}}^{v_{max}} \frac{1}{v_i}.
\end{equation}

Early work in this field \citep*{doi:10.1146/annurev.ns.38.120188.003535,LEWIN199687, Freese_2013} implemented the Standard Halo Model and the resulting Maxwell Boltzmann distribution.
However, Equation \ref{eq:rate_sum} allows any input velocity distribution to be used with no prior assumption of a Maxwellian form. This is achieved through the independent re-derivation of key coefficients $R_0$, $E_0$ and $<v>$. Where $R_0 = \int_{v_E = 0} dR = \frac{N_0 \sigma}{A}\int v dn = \frac{N_0 \sigma n_0}{A} <v>$. $\ E_0 = \frac{1}{2}M_D <v>^2 \frac{\pi}{4}$. The $\frac{\pi}{4}$ factor is needed for consistency with \cite{LEWIN199687} definition, $\ <v> = \frac{2}{\sqrt{\pi}}v_0$. This novel process enables simulation outputs to be fed directly into a simple set of equations to generate realistic predictions of the differential count rate of dark matter particles, as seen from Earth. The \textit{Dark MaRK} package utilizes this form of the equation.

We now demonstrate that Equation \ref{eq:dRdEr_derive} agrees with equation 3.9 in \cite{LEWIN199687} for the case of a truncated Maxwell Boltzmann velocity
distribution
\begin{equation}
  f_{geo}(\vec{v},\vec{v}_E) = \begin{cases}
    e^{-(\vec{v} + \vec{v}_E)^2/v_0^2} & v < v_{esc} \\
    0 & v > v_{esc}
\end{cases}.
\end{equation}
We define $R_0$, the total event rate per unit mass for $v_E = 0$ and
$v_{esc} = \infty$
\begin{equation}
  R_0 = \int dR = \frac{N_0 \, n_0 \, \sigma}{A \, k} \int_0^\infty v \, e^{-v^2/v_0^2} \, 4\pi v^2 dv = \frac{2 \, N_0 \, n_0 \, \sigma \, v_0}{A \, \pi^{1/2}}\,,
\end{equation}
and $E_0 = \frac{1}{2} M_D v_0^2$ as the most probable incident
kinetic energy, and $k_0 = \pi^{3/2} v_0^3$ as the value of $k$ for
$v_{esc} = \infty$.  In terms of these variables, the coefficient
outside Equation \ref{eq:dRdEr_derive} becomes
\begin{equation}
\frac{2 \, N_0 \, n_0 \, \sigma}{A \, k \, M_D \, r} = \frac{R_0 \,
  \pi^{1/2}}{v_0} \frac{1}{k \, r} \frac{v_0^2}{2 \, E_0}
\frac{k_0}{\pi^{3/2} v_0^3} = \frac{R_0}{E_0 \, r} \frac{k_0}{k}
\frac{1}{2\pi v_0^2}\,,
\end{equation}
agreeing with \cite{LEWIN199687} equation.

\subsection{Quenching Effect}
The quenching factor, $Q(E_R)$, is a function used to describe the conversion of nuclear recoil energies into `electron equivalent energies' as the energy detectable from a crystal detector, or simply the energy of the scintillation event
\begin{equation}\label{eqn:recoil}
E_{ee} = Q(E_R) \, E_R.
\end{equation}
A nuclear recoil can be distinguished from an electron recoil by observing the fraction of deposited energy released as scintillation. 
These signals, detected in the photo-multiplier tubes of direct detection experiments, are measured in `electron equivalent energies', \kevee \citep{doi:10.1146/annurev.nucl.54.070103.181244}.
This allows them to be used as a tool to discern WIMP recoils, which deposit energy via nuclear recoil, from background sources (primarily high-energy gamma and X-rays), which deposit energy via electron recoil.

This means that in order to find the observable differential event rate, for visible recoils, $\frac{dR}{dE_{ee}}$, the annual modulation integral needs to account for this relative efficiency. 
For an event rate detected in a given observational energy window (i.e. `ee') or band, $R_{\rm band}$, we average over that observational window ($\Delta E_{ee} = E_{ee}^{max}-E_{ee}^{min}$) as
\begin{equation}
\begin{split}
    \frac{ R_{\rm band}}{\Delta E_{ee}} &= \frac{1}{\Delta E_{ee}}\int_{E_{ee}^{min}}^{E_{ee}^{max}} \frac{dR}{dE_{ee}} dE_{ee} \\
    &=\frac{1}{(E_{ee}^{max}-E_{ee}^{min})} \int_{E_{R}^{min}}^{E_{R}^{max}} \frac{dR}{dE_R}  dE_{\rm R}.
    \label{eq:rband}
\end{split}
\end{equation}

For Germanium detectors, we follow the Lindhard formalism from \cite{Benoit_2007} and \cite{BARKER20138}
\begin{equation}\label{eqn:Ge_quench}
    Q = \frac{kg(\epsilon)}{1+kg(\epsilon)}\,,
\end{equation}
where
\begin{equation}
    g(\epsilon) = 3\epsilon^{0.15}+0.7\epsilon^{0.6}+\epsilon\,,
\end{equation}
and
\begin{equation}
   \epsilon = 11.5Z^{-\frac{7}{3}}E_{R}.
\end{equation}
Here $Z$ is the atomic number of the recoiling nucleus, $\epsilon$ a dimensionless energy, $E_{R}$ is the recoil energy in keV and $k$ describes the electronic energy loss.
We adopt the free electronic energy loss $k = 0.179\pm 0.001$ from \cite{Scholz_2016}.

For Sodium and Iodine, the light-yield ratio of the nuclear recoil to electron recoil is measured to be between 10-23\% (Q = 0.1-0.23) for Na in the energy range of 9-152keV. For I, the quenching range is 4-6\% within 19-75keV \citep{JOO201950}. \cite{Simon:2002cw} finds slightly higher Na quenching values of $25.4-29.4\%$ for $50-336$\keV. 

Here, we take the conservative scalar approximation of $Q(Na) = 0.3$ and $Q(I) = 0.09$, as adopted by the DAMA collaboration \citep{1996PhLB..389..757B}. This does not account for any energy dependence of the quenching factor. We combine these terms as described in Section \ref{sec:Results} below.

\subsection{Form Factor}\label{sec:Form_factor}
To account for the fact that simple scattering is not an appropriate way to model the interaction of large target nuclei with heavy WIMP dark matter, a model for nuclear charge density is introduced into dark matter detection rate calculations. This important factor explains the crucial velocity dependence of the interaction cross-section in Equation \ref{eq:dR}. This work implements the Woods-Saxon nuclear form factor for scalar interactions (a more accurate model than the Helm ansatz, as in \citealt{LEWIN199687})
\begin{equation}
    F(E_R)^2 = [\frac{3j_1(qr_1)}{qr_1}]^2 exp[-(qs^2)].
\end{equation}
This effective interaction of two nuclei undergoing an elastic collision is quantified at non-zero momentum transfer $q = \frac{\sqrt{2M_TE_R}}{\hbar c}$. The effective nuclear radius is $r_1 = \sqrt{r^2-5s^2}$ where we approximate $r = 1.2 \, {\rm fm} \times M_T^{\frac{1}{3}}$, and the nuclear skin thickness is $s \approx 1 \, {\rm fm}$ \citep{Jungman_1996}.

At zero momentum transfer, the effective cross-section then becomes 
\begin{equation}
\sigma(x) = \sigma_0F^2(x). \label{eqn:sigma}\,,
\end{equation}
where $\sigma_0$ is evaluated using Equation \ref{eq:sigma}.
The Form Factor thus acts to truncate high-energy recoil events. 
For further discussion of detector considerations for the SABRE experiment (i.e efficiency, resolution, sensitivity), we refer readers to \cite{Zurowski_2020}.

\subsection{Dark Matter candidate selection}
For ease of comparison we consider only two dark matter candidates in the 1-100\gevc range, following \cite{2008EPJC...56..333B}. 
The Low Mass Model (LMM) assumes a dark matter mass of 15\gevc and a cross section per nucleon of $\sigma_0=1.3\times10^{-41}$\cms. 
The High Mass Model (HMM) in Section \ref{sec:Ge_HMM} assumes a dark matter mass of 60\gevc and a cross section per nucleon of $\sigma_0=5.5\times10^{-42}$\cms. 
These are chosen from possible models outlined in \cite{2008EPJC...56..333B} for their alignment with Germanium and Sodium masses, 67.66\gevc and 21.44\gevc respectively.

Results are provided in two categories; evaluated for electron equivalent energies in Section \ref{sec:Results}, and nuclear recoil energies in Appendix \ref{app:E_NR}. The electron equivalent energies are observable by detectors as explored below and 2-6\kevee is the region of interest for DAMA and SABRE. Comparing these with nuclear recoil results, we can see the effects of quenching on reducing the emitted energy.

\section{Results}\label{sec:Results}
We now explore the impact of this more realistic, messier dark matter halo on the annual modulation signal from different detector types, first for Germanium (Section \ref{sec:Ge}) and then Sodium Iodide (Section \ref{sec:NaI}) detectors. We'll consider energy ranges defined by both nuclear recoil energies and electron equivalent energies.

\subsection{Germanium}\label{sec:Ge}
\subsubsection{Ge - Low Mass Model}\label{sec:Ge_LMM}

\begin{figure}
    \centering    \includegraphics[width=1.0\linewidth]{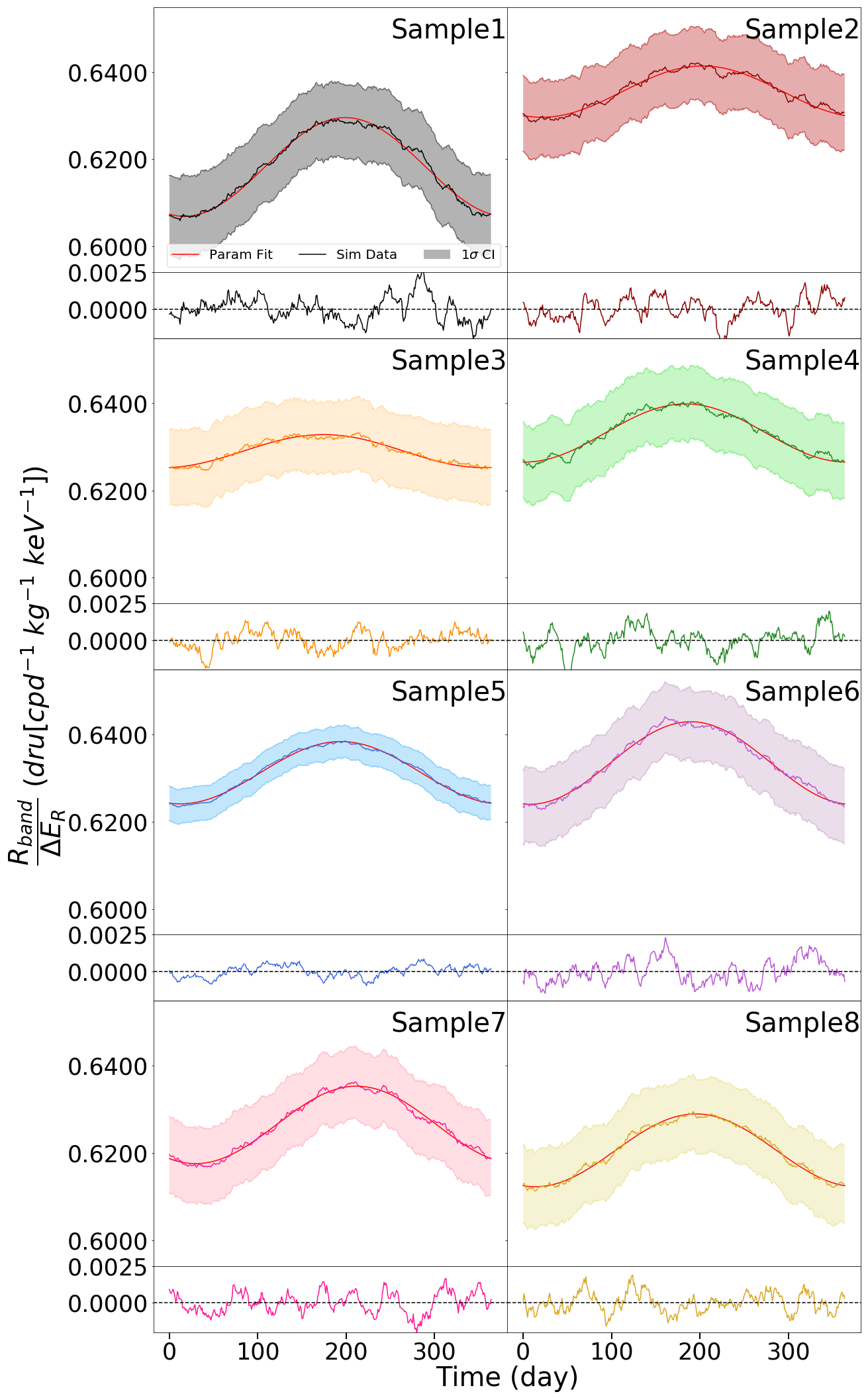}
    \caption{Annual modulation curves for each Solar Circle sample, evaluated per nucleon for the low-mass dark matter model for Germanium detectors between 2-6\keV (nuclear recoil energy, $E_R$). The 1$\sigma$ confidence intervals are estimated by bootstrap resampling. The event rate through the year is shown as a solid line, with the best-fitting annual modulation curve given by Equation \ref{eq:DAMA_fit} in red. Day 0 corresponds to January 1st of 2010 (`J2010’ equinox). Below each annual modulation subplot is a residual showing the difference between the simulation data and the parameter fit.}
    \label{fig:am_Ge_lowmass_2-6keV}
\end{figure}

Assuming a Germanium detector with nuclear mass 70 \amu\ and using Equations \ref{eq:rate_sum} \& \ref{eq:rband}, the annual modulation curves are evaluated for the energy bin 2-6 \keV and plotted in Figure \ref{fig:am_Ge_lowmass_2-6keV}. The shaded regions indicate 1$\sigma$ confidence intervals evaluated using a bootstrap resampling technique.
These samples demonstrate the typical sinusoidal shape of the annual modulation, peaking in the middle of the year. The red curve in Figure \ref{fig:am_Ge_lowmass_2-6keV} shows the best fit of the sinusoidal model of Equation \ref{eq:DAMA_fit} to the data.

We see some variation in the signal between samples, indicative of different velocity structures around the Solar Circle, together with noise arising from the number of particles.  Qualitatively, this creates visible changes in the signals between samples owing to these astrophysical effects.  Our sinusoidal fits allow us to explore the effect of this variation on summary statistics such as the peak day.

For the case of evaluating the rates at nuclear recoil energies (i.e. in \keV units), $S_m$ values range between $3.794\times 10^{-3} - 1.131\times 10^{-2}$ \dru\ (where \dru\ units are counts \counts), with a modulation fraction, $\frac{S_m}{S_0}$, of 0.60\%-1.83\%.

\begin{figure}
    \centering
    \includegraphics[width=1.0\linewidth]{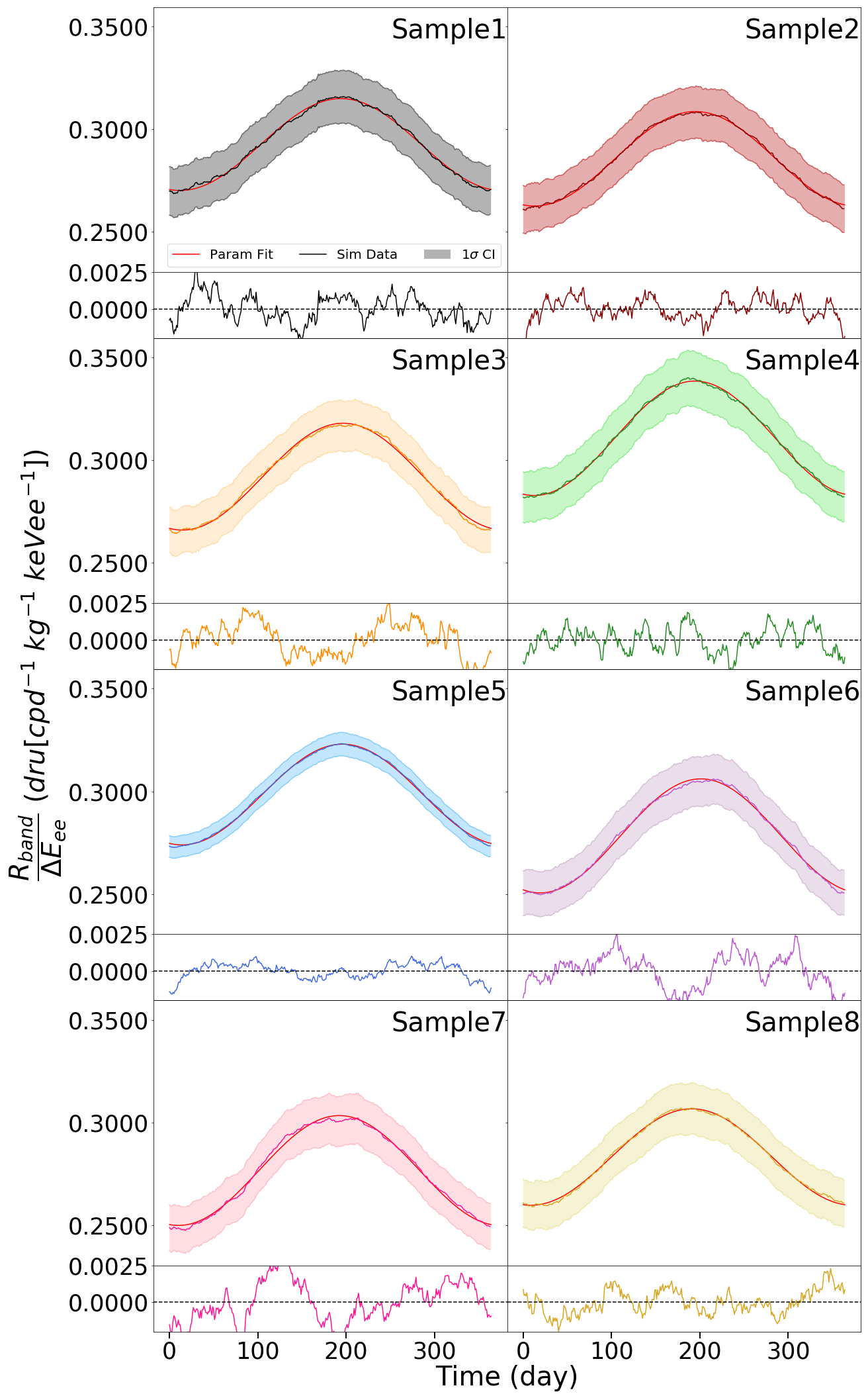}
    \caption{Similar to Figure \ref{fig:am_Ge_lowmass_2-6keV}, the annual modulation curves for each sample, evaluated per nucleon for the Low Mass Model dark matter interacting with Germanium, now considering a range of observed electron equivalent energies between 2-6\keVee. Shaded regions show 1$\sigma$ confidence intervals estimated from bootstrap resampling. Below each annual modulation subplot is a residual showing the difference between the simulation data and the parameter fit.}
    \label{fig:am_Ge_lowmass_2-6keVee}
\end{figure}

Re-evaluating this result in terms of electron-equivalent energy (observable by detectors), that quench via Equation \ref{eqn:Ge_quench}, we examine the energy region of 2-6\keVee.  In Figure \ref{fig:am_Ge_lowmass_2-6keVee}, we see the same general trends of a sinusoidal signal, peaking during the middle of the year and in agreement with each other within 1$\sigma$ confidence intervals.  We see values for $S_m$ ranging from $2.237\times 10^{-2}-2.786\times 10^{-2}$\dru\ with fractional modulations of 7.65\%-9.97\%.

\begin{figure}
    \centering
    \includegraphics[width = 1.1\linewidth]{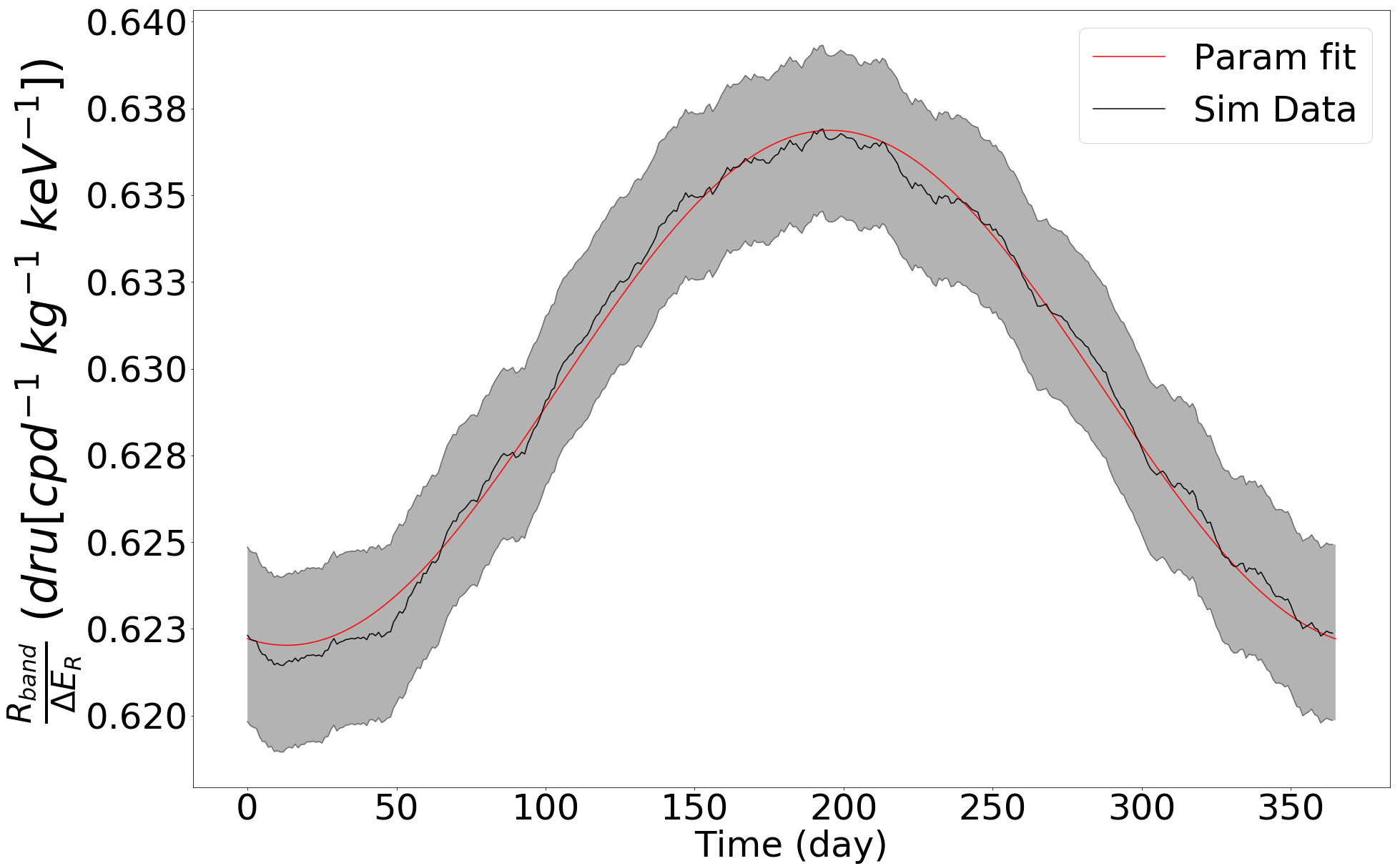}
    \caption{Annual modulation curve for the Solar Circle sample, evaluated per nucleon for the Low Mass Model dark matter interacting with Germanium between 2-6\keV. The shaded region demonstrates$\ 1\sigma$ confidence intervals.}
    \label{fig:am_Ge_lowmass_2-6keV_tot}
\end{figure}

\begin{figure}
    \centering
    \includegraphics[width = 1.1\linewidth]{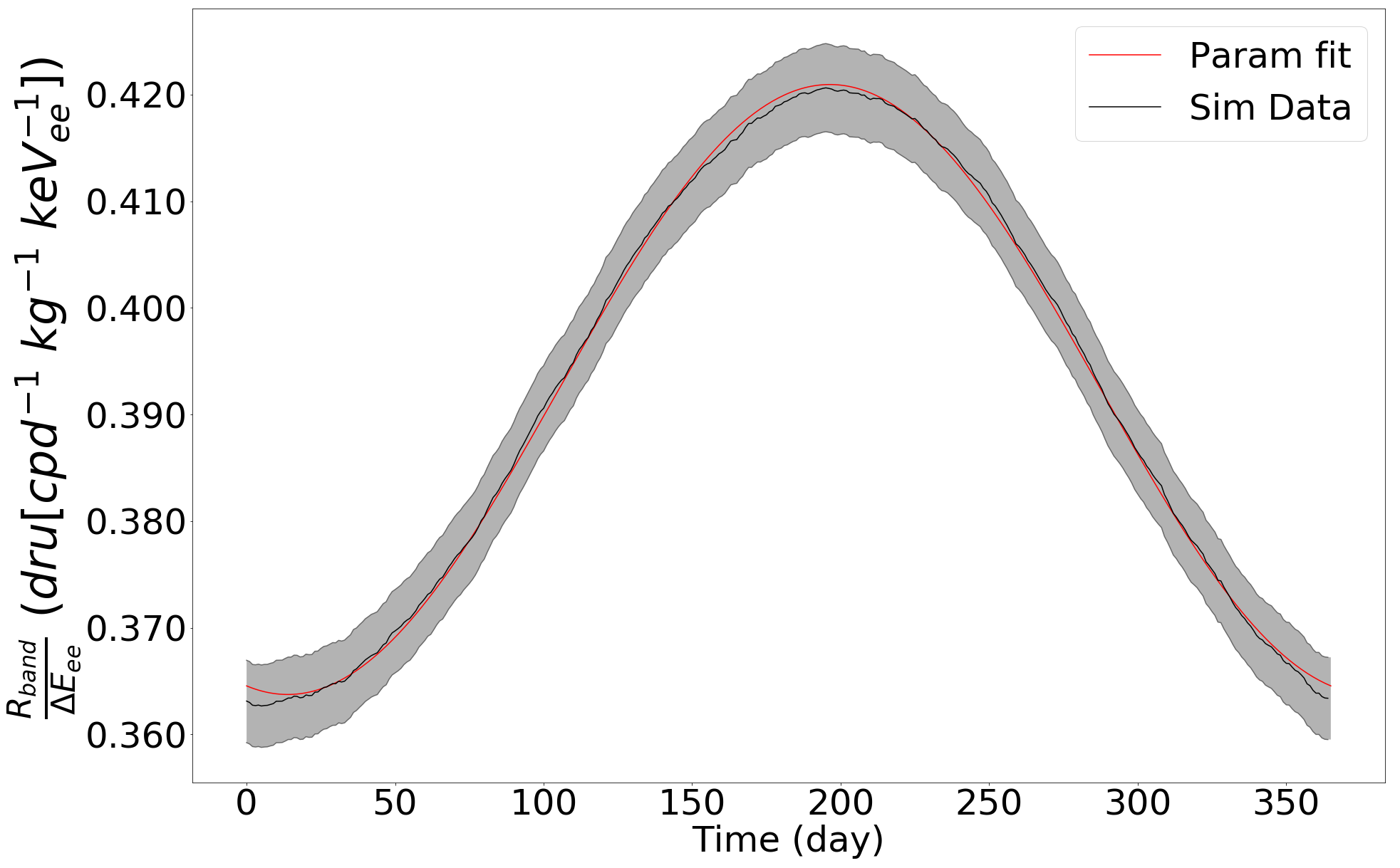}
    \caption{Annual modulation curve for the Solar Circle samples, evaluated per nucleon the Low Mass Model dark matter interacting with Germanium between 2-6\keVee. The shaded region demonstrates$\ 1\sigma$ confidence intervals.}
    \label{fig:am_Ge_lowmass_2-6keVee_tot}
\end{figure}

When looking at all of the samples combined, at a `Solar Circle' perspective, we can gain a better idea of the underlying mean distributions. By stacking the particle samples in this way,  discreteness effects due to particle counts within a sample are reduced. Additionally, we smooth out the fluctuations observed between the samples. Results of such an analysis are shown in Figures \ref{fig:am_Ge_lowmass_2-6keV_tot} and \ref{fig:am_Ge_lowmass_2-6keVee_tot}.

The difference between Figures \ref{fig:am_Ge_lowmass_2-6keV_tot} and \ref{fig:am_Ge_lowmass_2-6keVee_tot} can be explained by the fact that $2-6$\keV is not equivalent to $2-6$\kevee. 
The region of experimental interest and sensitivity, $2-6$\kevee for Germanium quenching values, corresponds to 9.67-24.9\keV. This emphasizes the importance of modelling our quenching factors accurately, and how their uncertainty can impact the interpretation of our observations. The larger energy range associated with 2-6\keVee allows for more counts to be contained within that region of interest, allowing for tighter constraints on the scatter within the signal.

\begin{figure}
    \centering
    \includegraphics[height=1.5\linewidth]{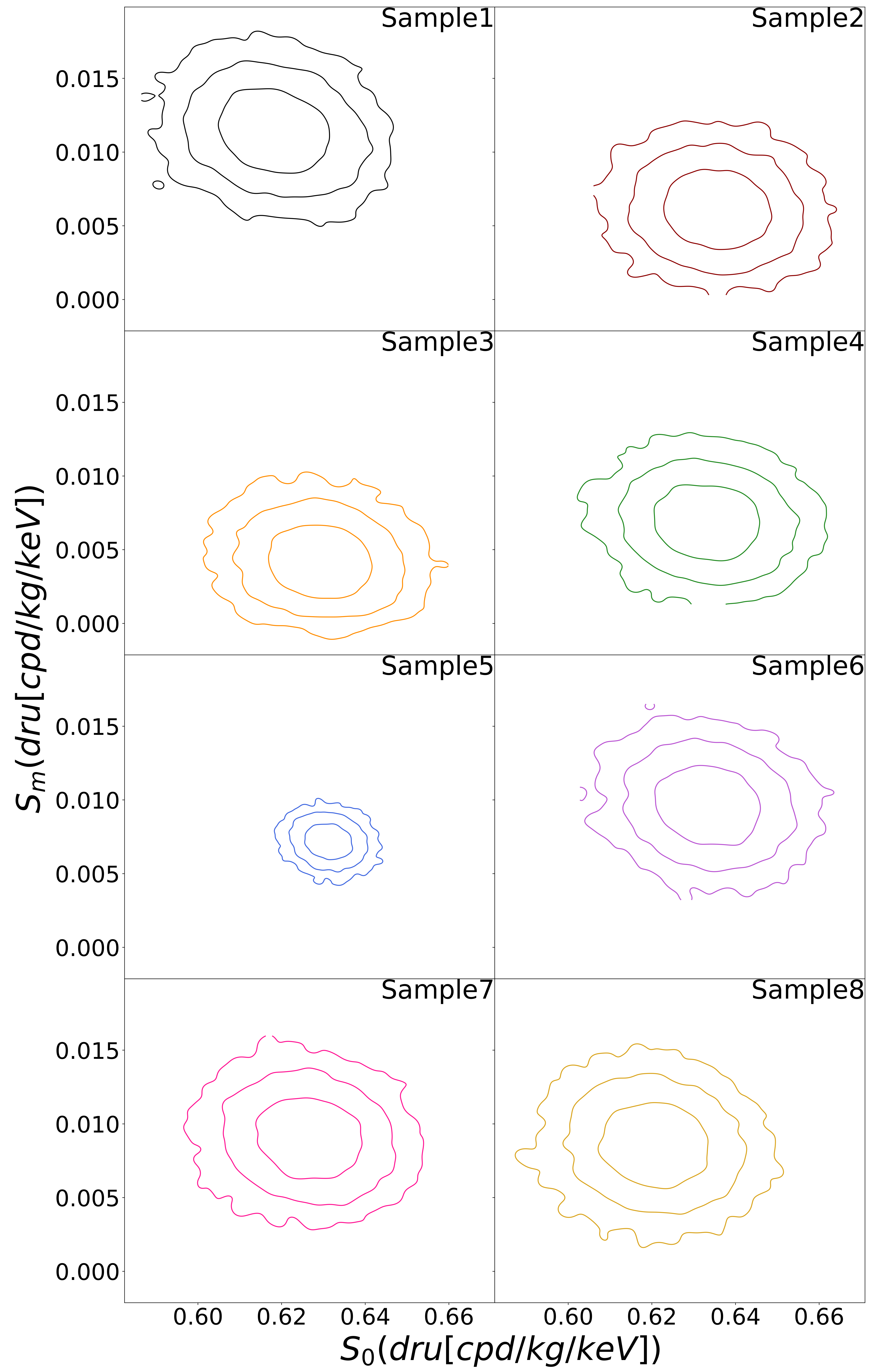}
    \caption{Total rate ($S_0$) versus amplitude ($S_m$) for a 15$GeVc^{-2}$ WIMP interacting with a Germanium detector, evaluated at 2-6\keV. Contours represent 1,2,3$\sigma$ confidence intervals.}
    \label{fig:S0_Sm_Ge_lowmass_keV}
\end{figure}

\begin{figure}
    \centering
    \includegraphics[height=1.5\linewidth]{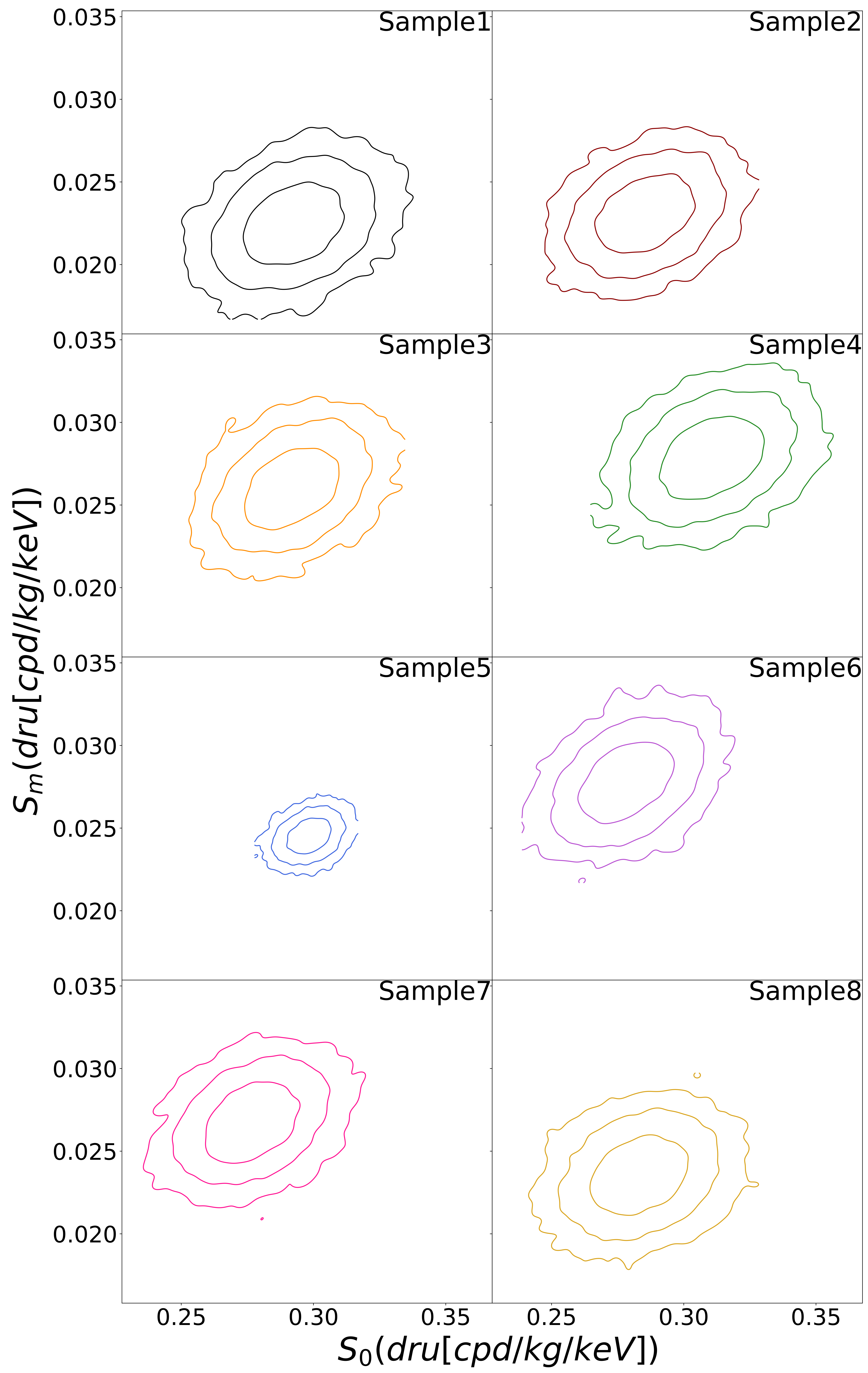}
    \caption{Total rate ($S_0$) versus amplitude ($S_m$) for a 15\gevc WIMP interacting with a Germanium detector, evaluated at 2-6\keVee. Contours from inside out represent 1,2,3$\sigma$ confidence intervals respectively.}
    \label{fig:S0_Sm_Ge_lowmass_keVee}
\end{figure}

Figures \ref{fig:S0_Sm_Ge_lowmass_keV} and \ref{fig:S0_Sm_Ge_lowmass_keVee} are joint confidence regions of the fit parameters $\ S_m, S_0, t_0$ to understand the errors and covariance in the parameters using all of our bootstrap resamples. This allows us to look for any correlations and better understand the variation  within, and between, samples in these fits.


There is no apparent correlation or degeneracy between the fitted parameters (as expected from theory) within a sample. The quality of the fits are however impacted by the location. For example, Sample 5 appears better constrained due to its higher particle density. 
The relationship between $S_m$ and $t_0$, and $S_0$ and $t_0$, follow the same narrative and their values are listed in Table \ref{table:Ge_fits}, with their plots available in the author's supplementary \href{https://github.com/Grace-Lawrence/Gusts_in_the_Headwind-Uncertainties_in_Direct_Dark_Matter_Detection-Supplementary}{repository}. The $S_0$ values are quoted for completeness, however we note that the modulation amplitude, $S_m$, is the experimentally significant parameter. 

\subsubsection{Peak Day}
We next explore the peak day of the detection rate fluctuations from each sample. Traditionally this would occur when the velocity of the Earth aligns with the Sun and `static' or completely randomly moving dark matter background. As we will see, the impact of inherent variations in the dark matter wind complicates this simple expectation. We note that these dates should be compared against the fiducial expectation value of $t_0 = 152.5$ (corresponding to June 2nd 2021) \citep{Bernabei_2018}. 

\begin{figure}
 \begin{center}
  \includegraphics[width = \linewidth]{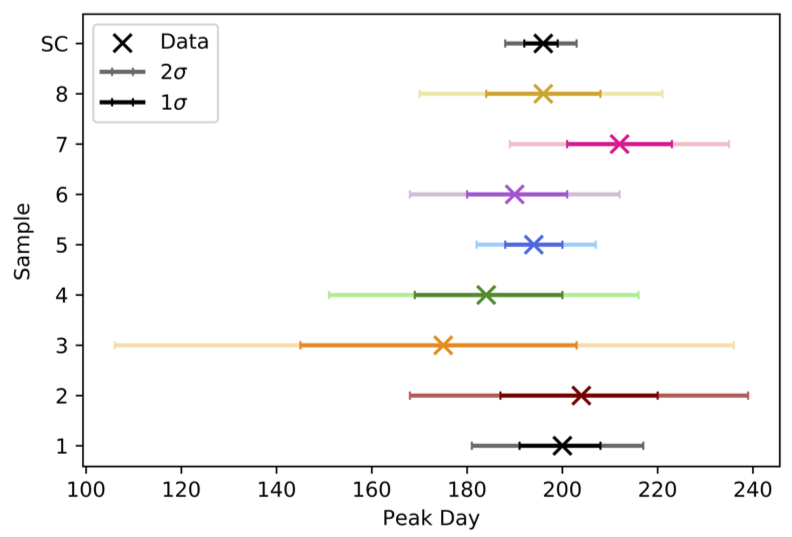}
  \includegraphics[width = \linewidth]{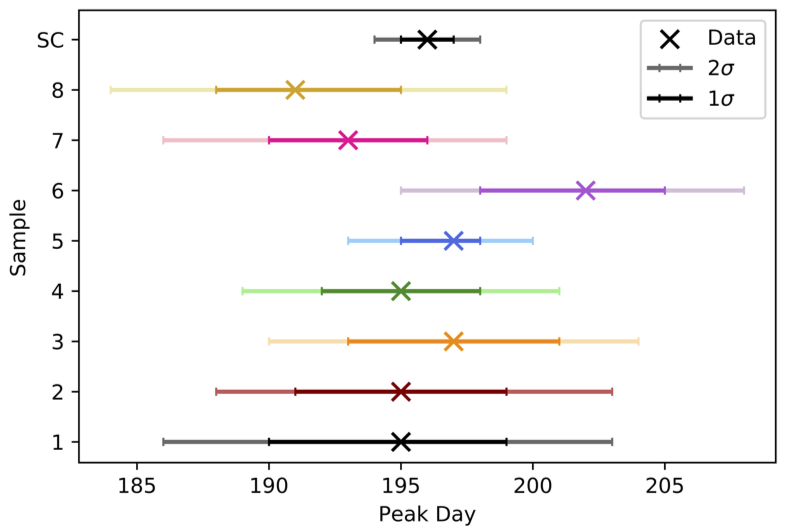}
\caption{In the top (bottom) panel we see the peak day of dark matter counts for LMM Germanium according to the annual modulation curves evaluated for nuclear recoils (electron equivalent energies) with 1,2$\sigma$ errors as estimated by bootstrap resampling. The `Data' cross is the day estimated using the full particle sample. The top data point represents the combined Solar Circle (SC) sample.}\label{fig:Ge_lowmass_peakday}
\end{center}
\end{figure} 

Figure \ref{fig:Ge_lowmass_peakday} shows our measurements of the peak days for each sample, which lie between day 174-212 for evaluations between 2-6\keV (nuclear recoil energy)  and day 191-201 for evaluations between 2-6\kevee  (electron equivalent energies). For the year 2021, these correspond to dates between June $23^{rd}$ and July $31^{st}$.

We find that the best-fitting peak days are statistically consistent across the different samples, although the size of the error can vary significantly between samples.

\subsubsection{Ge - High Mass Model}\label{sec:Ge_HMM}
For the High Mass Model (HMM) (as in Section \ref{sec:Ge_LMM}) we evaluate the annual modulation predictions for electron equivalent energy regions of 2-6 \kevee. The nuclear recoil energy evaluations are listed in Appendix \ref{app:E_NR} and Table \ref{table:Ge_fits} for reference. 

\begin{figure}
    \centering
    \includegraphics[width=1.0\linewidth]{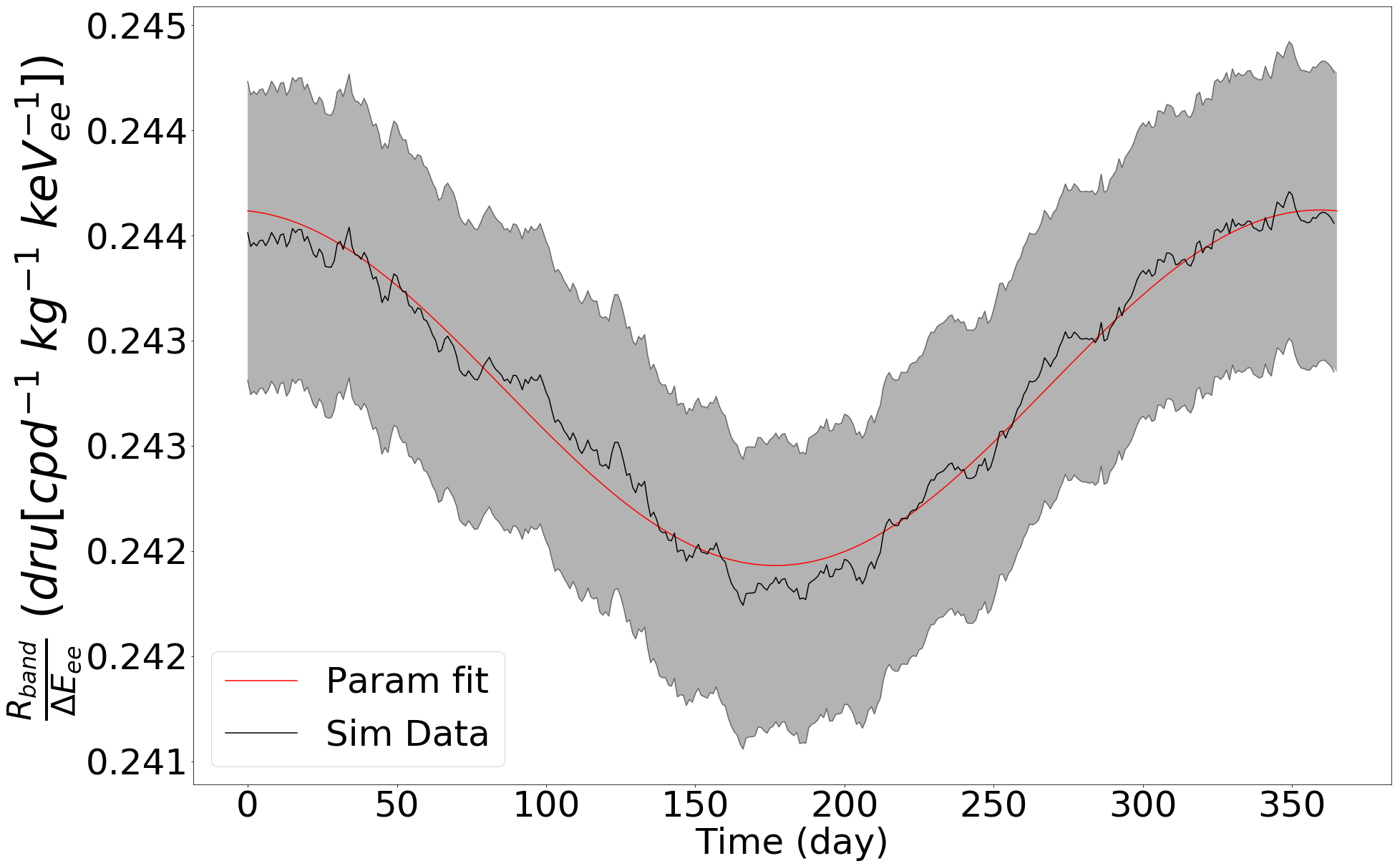}
    \caption{Annual Modulation curve for the combined total of all particles sampled, evaluated per nucleon for HMM Germanium between 2-6\kevee. 1$\sigma$ confidence intervals are shown.}
    \label{fig:am_Ge_highmass_2-6keVee_tot}
\end{figure}

The narrative is very similar to that of the LMM and we refer readers to the \supp~to view sample-specific results. 
We present, as an overview, the Solar Circle plots in Figure \ref{fig:am_Ge_highmass_2-6keVee_tot} and Figure \ref{fig:am_Ge_highmass_2-6_tot}. The rate here undergoes a few key changes with the different dark matter mass model. 
The annual modulation, $S_m$, of the nuclear recoil signal has decreased by approximately 3.6 times and the electron equivalent energy by 57.2 times. 
The signal, while still sinusoidal, has also undergone a 180$^\circ$ phase shift. This phase flip occurs at low recoil energies ($\propto v_{min}$) where the phase shift for different recoil energies can change depending on the assumed dark matter mass. 

\begin{figure}
    \centering
    \includegraphics[width=86mm]{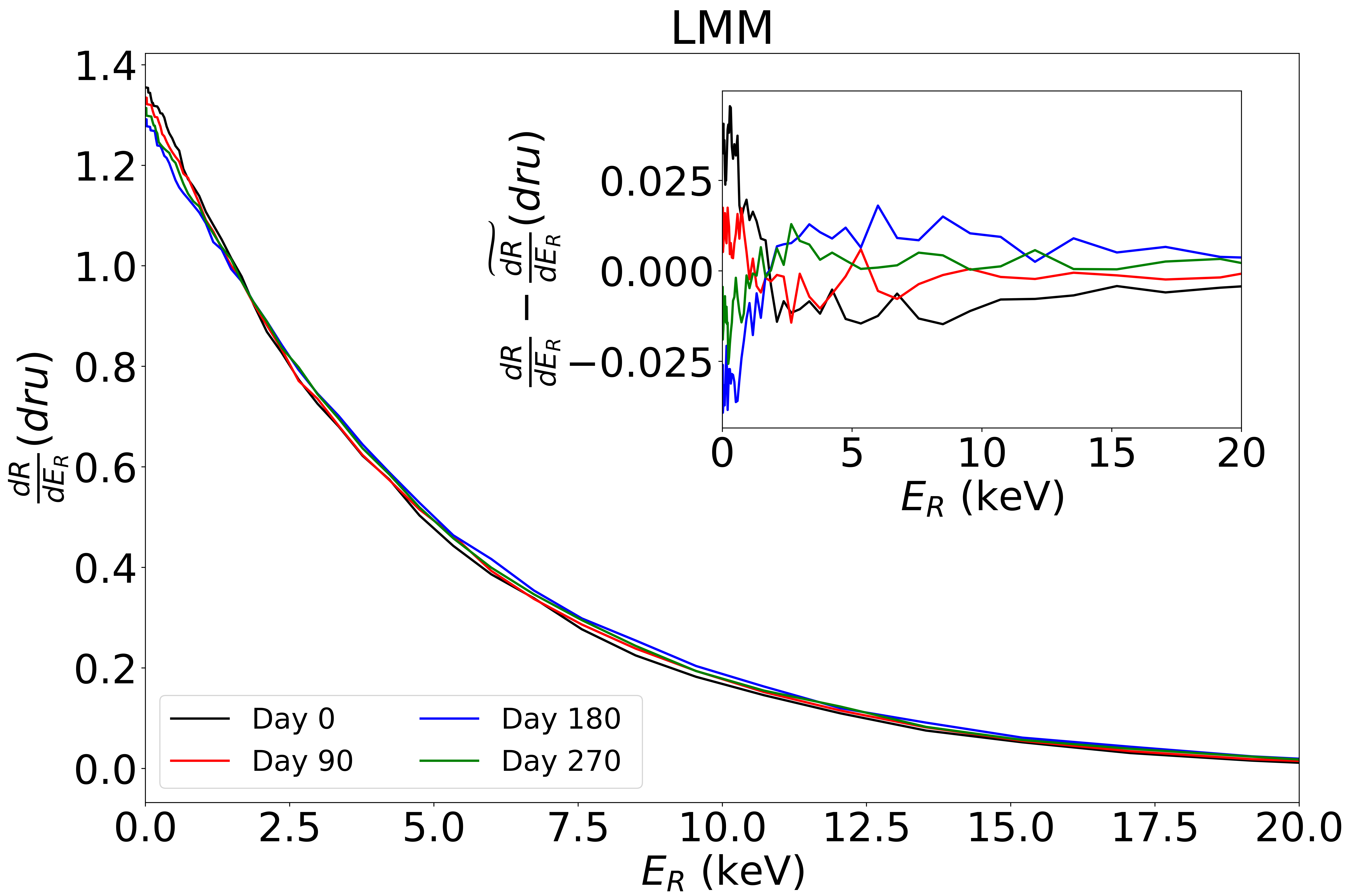}
    \caption{The spectral function for the LMM Germanium detector in dru units. The colored lines represent four evenly spaced times during the year. The inset shows the same plot, with the annual average subtracted from the time samples, and with energy on the x-axis. }
    \label{fig:Ge_LMM_specfunc}
\end{figure}

\begin{figure}
    \centering
    \includegraphics[width=85mm]{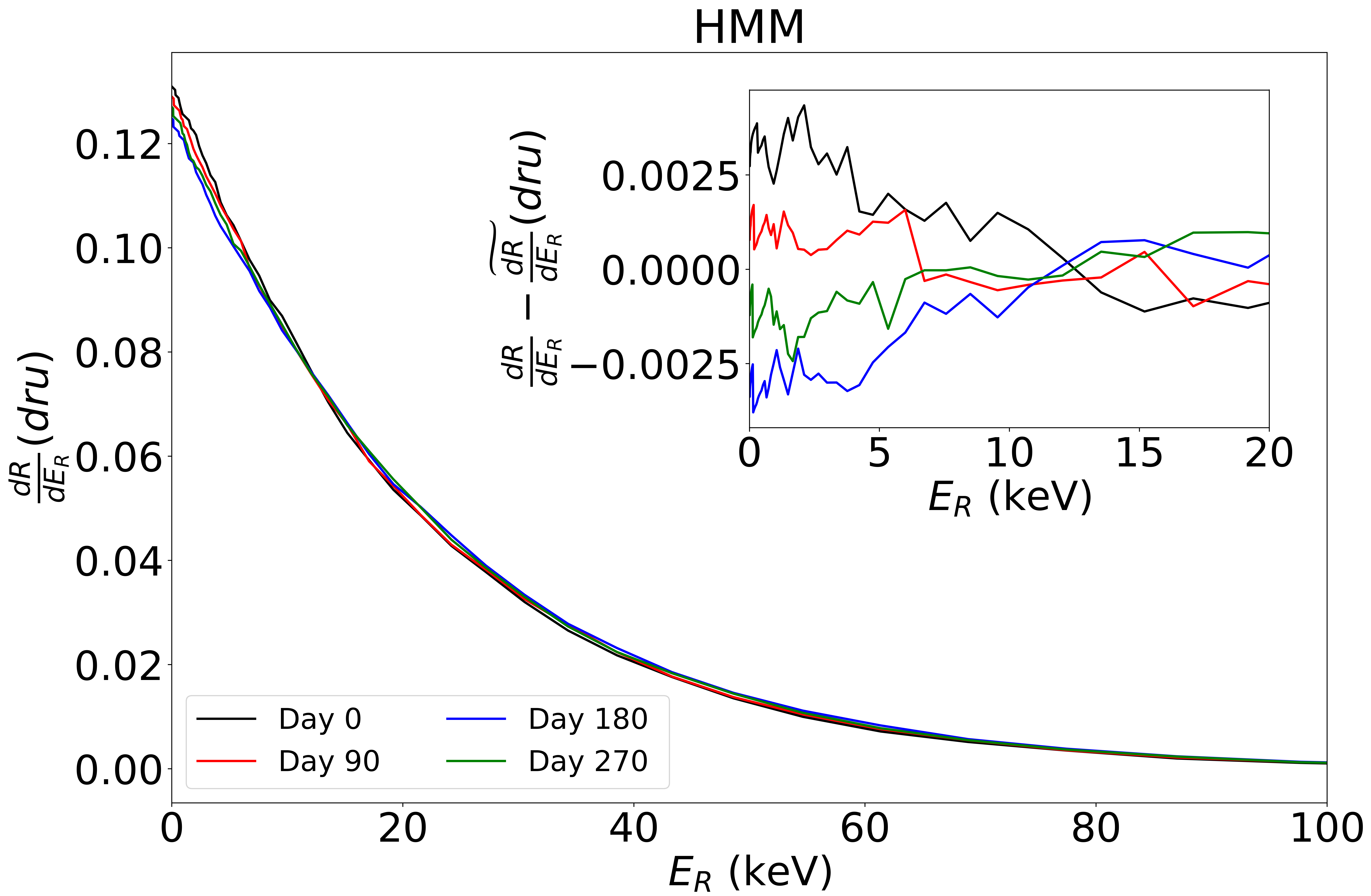}
    \caption{The spectral function for the HMM Germanium detector. The colored lines represent four evenly spaced times during the year. The inset shows the same plot, with the annual average subtracted from the time samples, and with energy on the x-axis. }
    \label{fig:Ge_HMM_specfunc}
\end{figure}

Best-fitting parameters for this version can be found in Table \ref{table:Ge_fits}.
We note that the higher mass WIMP, with correspondingly smaller cross-section, undergoes a phase inversion within the 2-6\keV/\kevee energy interval.
The 60\gevc dark matter particle strongly couples kinematically with the 65.24\gevc (70amu) Germanium nucleus. 
This effect influences the rate through the $\frac{1}{AM_D^2}$ factor in the coefficients of Equation \ref{eq:rate_sum}, causing spectral functions of the HMM to be lower than those of the LMM. 
The spectral functions, found in Figures \ref{fig:Ge_LMM_specfunc}-\ref{fig:Ge_HMM_specfunc}, demonstrate the seasonal variation in rate, where their point of intersection is indicative of the phase. 
Comparison of these rates shows that the turnover will occur at different energies, for different dark matter particle candidates. 


The phase inversion at some critical energy $Q_c$, determined by the dark matter mass, occurs at low recoil energies (or high $v_{min}$ values).
The phase of the modulation is fixed for a given $v_{min}$, however the phase of the modulation for a given recoil energy is not \citep{Freese_2013}. 
By evaluating the spectrum at different recoil energies, the phase will change. 
This phase flip is not only an inherent feature of dark matter detectors with serious ramifications for the interpretation of experimental results, but also a feature which is highly sensitive to the uncertainties of the astrophysical input parameters. 
This will be explored fully in our subsequent work, exploring the constraints that can be put on dark matter particle mass in Germanium and Sodium Iodide detectors by taking advantage of this phenomena.

\subsubsection{Peakday}
\begin{figure}
   \centering
 \includegraphics[width=\linewidth]{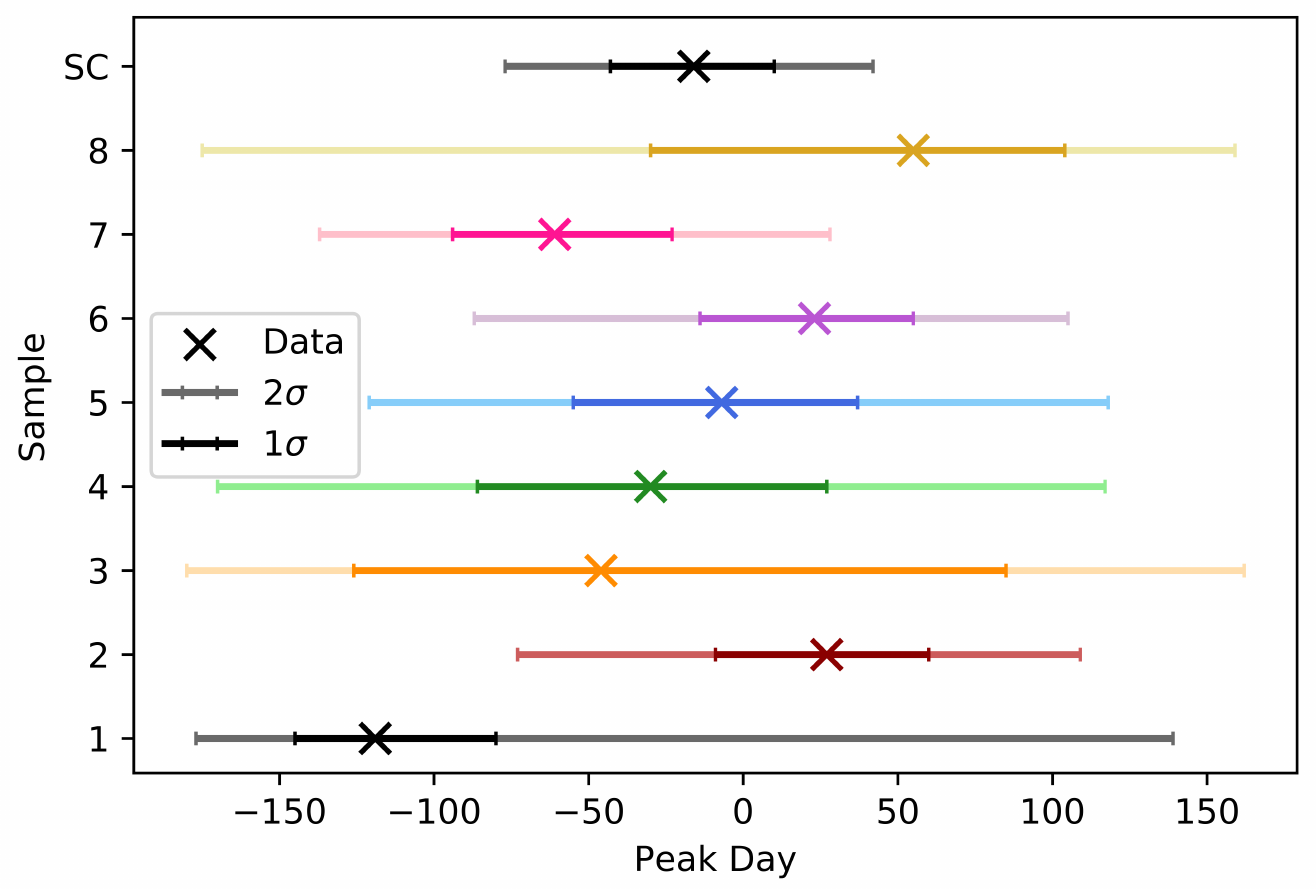}
     \caption{Peak day of dark matter counts according to the annual modulation curves of the HMM Germanium evaluated for electron equivalent energies with 1,2$\sigma$ confidence intervals.}
     \label{fig:Ge_highmass_peakday_Eequiv}
 \end{figure}
 
The phase shift in the HMM moves the expected peak flux of dark matter from the middle of the year, to the end. Peak days for \kevee considerations range from day 245-55. These correspond to dates between September 2$^{nd}$ and February 24$^{th}$. The standard deviation between samples is 144 days. 
  
Compared to the 2-6\keV range, the 2-6\kevee range has a much broader error distribution. This large error range arises from the fact that 2-6\keVee, for the HMM 60\gevc model for a Germanium detector, corresponds closely to approximations for $Q_c$ for Germanium. Estimated to fall within the 2-6\kevee energy regime \citep{Lewis_2004}, a precise calculation of $Q_c$ for a given detector, incorporating the realistic messy halo from this research, will be explored in greater detail in future work (Lawrence et al. in prep).

The annual modulation amplitude, $S_m$, is negligible so the peak day is not well constrained. This indicates that the inherent variation around the Solar Circle may push $t_0$ to be slightly below or above $Q_c$, causing a subsequent error range that comprises over $\sim 50$\% of the sample. 

\begin{landscape}
\begin{table}
\centering
\begin{tabular}{|l|l|l|l|l|l|l|l|l|l|} 
            & \multicolumn{3}{|c|}{\textbf{Nuclear Energy $E_{R}$ - 2-6\keV}} &\multicolumn{3}{|c|}{\textbf{Electron Equivalent Energy $E_{ee}$ - 2-6\kevee}} &\multicolumn{3}{|c|}{\textbf{Electron Equivalent Energy MB Fit $E_{ee}$ - 2-6\kevee}} \\
            & {\textbf{$S_0$ [dru]}}     & {\textbf{$S_m$[mdru]}}    & {\textbf{$t_0$ [day]}} & {\textbf{$S_0$ [dru]}}     & {\textbf{$S_m$[mdru]}}    & {\textbf{$t_0$ [day]}} & {\textbf{$S_0$ [dru]}}     & {\textbf{$S_m$[mdru]}}    & {\textbf{$t_0$ [day]}} \\ \hline
\textbf{S1} &	$0.618_{0.017}^{0.017}$	&	$11.314_{3.512}^{3.798}$ &	$199.636_{18.785}^{17.697}$ & $0.292_{0.012}^{0.012}$ &	$22.368_{1.518}^{1.670}$ & $194.627_{4.220}^{4.167}$ & $0.307_{0.012}^{0.012}$	&	$23.614_{1.608}^{1.695}$ &	$194.578_{3.950}^{3.788}$\\ \hline
\textbf{S2} &	$0.636_{0.016}^{0.016}$	&	$5.883_{3.167}^{3.753}$  &	$203.524_{35.373}^{35.534}$ & $0.286_{0.012}^{0.012}$	&	$23.001_{1.464}^{1.606}$	&	$195.340_{3.842}^{3.912}$ &	$0.275_{0.012}^{0.012}$	&	$22.534_{1.479}^{1.582}$ &	$197.827_{4.054}^{3.951}$\\ \hline
\textbf{S3} &	$0.629_{0.016}^{0.016}$	&	$3.794_{2.612}^{3.769}$  &	$174.900_{68.202}^{62.722}$ &$0.292_{0.012}^{0.012}$	&	$26.002_{1.555}^{1.603}$	&	$197.058_{3.720}^{3.564}$ &	
$0.284_{0.012}^{0.012}$	&	$23.486_{1.516}^{1.643}$	&	$200.523_{3.941}^{3.890}$\\ \hline
\textbf{S4} &	$0.633_{0.017}^{0.016}$	&	$6.669_{3.160}^{3.540}$  &	$184.261_{33.496}^{31.657}$ &	$0.311_{0.013}^{0.013}$	&	$27.857_{1.606}^{1.633}$	&	$194.944_{3.082}^{3.061}$ &	
$0.272_{0.012}^{0.012}$	&	$22.424_{1.484}^{1.595}$ &	$198.162_{3.781}^{3.653}$\\\hline
\textbf{S5} &	$0.631_{0.007}^{0.007}$	&	$7.147_{1.534}^{1.625}$  &	$194.473_{12.616}^{12.291}$ &	$0.298_{0.006}^{0.005}$	&	$24.522_{0.702}^{0.705}$	&	$196.521_{1.634}^{1.694}$ &	
$0.279_{0.005}^{0.005}$	&	$23.604_{0.697}^{0.692}$ &	$197.105_{1.677}^{1.641}$\\\hline
\textbf{S6} &	$0.633_{0.016}^{0.016}$	&	$9.467_{3.326}^{3.736}$  &	$190.151_{22.507}^{21.626}$ &	$0.278_{0.011}^{0.012}$	&	$27.758_{1.504}^{1.601}$	&	$201.629_{3.291}^{3.277}$ &	
$0.238_{0.010}^{0.011}$	&	$24.534_{1.524}^{1.657}$ &	$194.410_{3.419}^{3.490}$\\\hline
\textbf{S7} &	$0.626_{0.017}^{0.016}$	&	$8.880_{3.410}^{3.762}$  &	$212.320_{23.371}^{22.659}$ &	$0.277_{0.012}^{0.012}$	&	$26.737_{1.527}^{1.625}$	&	$192.763_{3.204}^{3.276}$ &	
$0.274_{0.012}^{0.012}$	&	$24.564_{1.601}^{1.674}$ &	$198.138_{3.434}^{3.565}$\\\hline
\textbf{S8} &	$0.621_{0.017}^{0.017}$	&	$8.346_{3.465}^{3.903}$  &	$196.114_{26.285}^{24.450}$ &	$0.283_{0.012}^{0.012}$	&	$23.477_{1.464}^{1.569}$	&	$191.260_{3.663}^{3.728}$ &	
$0.250_{0.011}^{0.011}$	&	$21.484_{1.520}^{1.670}$ &	$196.564_{4.394}^{4.246}$\\\hline
\textbf{SC} &	$0.629_{0.002}^{0.002}$	&	$7.421_{0.496}^{0.528}$  &	$195.581_{3.951}^{3.803}$ & $0.293_{0.003}^{0.003}$	&	$24.970_{0.453}^{0.459}$  &	$195.938_{1.047}^{1.053}$ &	$0.275_{0.003}^{0.003}$ &	$23.571_{0.455}^{0.467}$ &	$196.797_{1.074}^{1.065}$ \\\hline

\end{tabular}
\end{table}
\begin{table}
\centering
\begin{tabular}{|l|l|l|l|l|l|l|l|l|l|} 
            & \multicolumn{3}{|c|}{\textbf{Nuclear Energy $E_{R}$ - 2-6\keV}} &\multicolumn{3}{|c|}{\textbf{Electron Equivalent Energy $E_{ee}$ - 2-6\kevee}} &\multicolumn{3}{|c|}{\textbf{Electron Equivalent Energy MB Fit $E_{ee}$ - 2-6\kevee}} \\
            & {\textbf{$S_0$ [dru]}}     & {\textbf{$S_m$[mdru]}}    & {\textbf{$t_0$ [day]}} & {\textbf{$S_0$ [dru]}}     & {\textbf{$S_m$[mdru]}}    & {\textbf{$t_0$ [day]}} & {\textbf{$S_0$ [dru]}}     & {\textbf{$S_m$[mdru]}}    & {\textbf{$t_0$ [day]}} \\ \hline
\textbf{S1} &	$0.108_{0.002}^{0.002}$	&	$2.750_{0.658}^{0.706}$	&	$16.849_{12.429}^{348.151}$ &	$0.223_{0.002}^{0.002}$	&	$1.298_{0.479}^{0.856}$	&	$245.999_{30.985}^{31.885}$ & $0.221_{0.002}^{0.002}$ & $	0.971_{0.323}^{0.837}$ & $264.418_{48.758}^{34.909}$ \\	\hline
\textbf{S2} &	$0.107_{0.002}^{0.002}$	&	$1.880_{0.660}^{0.676}$	&	$14.576_{11.568}^{350.424}$ &	$0.225_{0.002}^{0.002}	$	&	$1.128_{0.395}^{0.781}$	&	$26.821_{12.941}^{288.125}$&$	0.227_{0.002}^{	0.002}$ & $	1.034_{0.333}^{0.807} $ & $	337.102_{315.466}^{10.119} $\\	\hline
\textbf{S3} &	$0.107_{0.002}^{0.002}$	&	$1.708_{0.649}^{0.709}$	&	$365.000_{364.396}^{0.000}$ &	$0.225_{0.002}^{0.002}	$	&	$0.203_{0.199}^{1.125}$	&	$318.596_{242.245}^{16.814}$&$	0.223_{0.002}^{	0.002}$ & $	0.497_{0.066}^{0.904} $ & $	348.222_{293.686}^{6.941} $\\	\hline
\textbf{S4} &	$0.106_{0.002}^{0.002}$	&	$1.522_{0.640}^{0.683}$	&	$27.420_{22.114}^{329.760}$ &	$0.224_{0.002}^{0.002}	$	&	$0.599_{0.093}^{0.969}$	&	$0.000_{7.056}^{330.687}$ &$	0.222_{0.002}^{	0.002}$ & $	0.302_{0.098}^{1.020} $ & $	359.660_{310.035}^{16.214} $\\	\hline
\textbf{S5} &	$0.107_{0.001}^{0.001}$	&	$1.927_{0.309}^{0.311}$	&	$19.372_{8.893}^{11.451}$ &	$0.223_{0.001}^{0.001}	$	&	$0.387_{0.123}^{0.374}$	&	$357.927_{321.237}^{2.820}$&$	0.224_{0.001}^{	0.001}$ & $	0.456_{0.155}^{0.368} $ & $	66.741_{27.289}^{61.814} $ \\	\hline
\textbf{S6} &	$0.108_{0.002}^{0.002}$	&	$2.472_{0.649}^{0.660}$	&	$15.727_{12.917}^{349.273}$ &	$0.228_{0.002}^{0.002}	$	&	$1.012_{0.343}^{0.857}$	&	$365.000_{327.143}^{0.000}$&$	0.223_{0.002}^{	0.002}$ & $	0.934_{0.320}^{0.840} $ & $	355.491_{342.370}^{7.214} $\\	\hline
\textbf{S7} &	$0.108_{0.002}^{0.002}$	&	$2.421_{0.604}^{0.641}$	&	$19.490_{16.288}^{17.627}$ &	$0.228_{0.002}^{0.002}	$	&	$1.123_{0.348}^{0.785}$	&	$304.482_{48.445}^{27.788}$&$	0.226_{0.002}^{	0.002}$ & $	0.533_{0.061}^{0.967} $ & $	298.960_{195.102}^{30.528} $\\	\hline
\textbf{S8} &	$0.109_{0.002}^{0.002}$	&	$2.089_{0.723}^{0.838}$	&	$365.000_{362.861}^{0.000}$ &	$0.225_{0.002}^{0.002}	$	&	$0.603_{0.136}^{0.912}$	&	$54.797_{33.954}^{147.552}$&$	0.224_{0.002}^{	0.002}$ & $	0.899_{0.283}^{0.904} $ & $	74.939_{36.242}^{63.246} $\\ \hline
\textbf{SC} &	$0.107_{0.000}^{0.000}$	&	$2.038_{0.095}^{0.102}$  &	$17.394_{2.813}^{2.830}$ & $0.224_{0.001}^{0.001}$	&	$0.436_{0.132}^{0.229}$  &	$349.314_{60.312}^{14.375}$ &	$0.222_{0.001}^{0.001}$ &	$1.086_{0.183}^{0.212}$ &	$33.220_{9.705}^{10.620}$ \\\hline
\end{tabular}
\centering
\caption{Parameter fits from Equation \ref{eq:DAMA_fit} with 1 $\sigma$ errors quoted. Classified for the LMM (top) and HMM (bottom) for a Germanium detector, and evaluated for (left) nuclear recoil energies, (centre) electron equivalent energies and (right) electron equivalent energies for Maxwell Boltzmann fits. $S_0$ values are expressed in dru units, $S_m $ in milli-dru units and $t_0$ in days.}\label{table:Ge_fits}
\end{table}
\end{landscape}

\subsection{Sodium Iodide}\label{sec:NaI}
Another widely-implemented detector compound is Sodium Iodide, which was used by the DAMA collaboration to provide their claim of evidence for annual modulation due to dark matter. The SABRE experiment will also use Sodium Iodide crystals implemented in a dual-hemisphere direct detection experiment \citep{Bignell_2020}.

Analysing Sodium Iodide requires an understanding of how the detector operates as a compound, including which element will provide dominant interactions with the dark matter. For our Sodium Iodide model, the Sodium and Iodine recoil energy spectrum are individually evaluated and then combined according to their relative weight ($f_{Na},~f_{I}$) and abundance ratio of the crystal (1:1 for NaI) according to
\begin{equation}\label{eq:weighted_fraction}
    \frac{dR}{dE_{ee}} = \sum_{\chi} f_{\chi} \left( \frac{dR}{dE_{ee}} \right)_{\chi} F_{\chi}^2 \,I_{\chi},
\end{equation}
where $f_{\chi} \equiv \frac{A_{\chi}}{(A_{Na}+A_{I})}$, $\left( \frac{dR}{dE_{ee}}\right)_{\chi}$ is the element specific spectral rate function, and $I_{\chi}$ is the given element $\chi$'s interaction term \citep{LEWIN199687}. The Sodium Iodide annual modulation curves implement Sodium's quenching factor, $Q(Na) = 0.3$.

\subsubsection{NaI - Low Mass Model}
\begin{figure}
     \centering
     \includegraphics[width=1.0\linewidth]{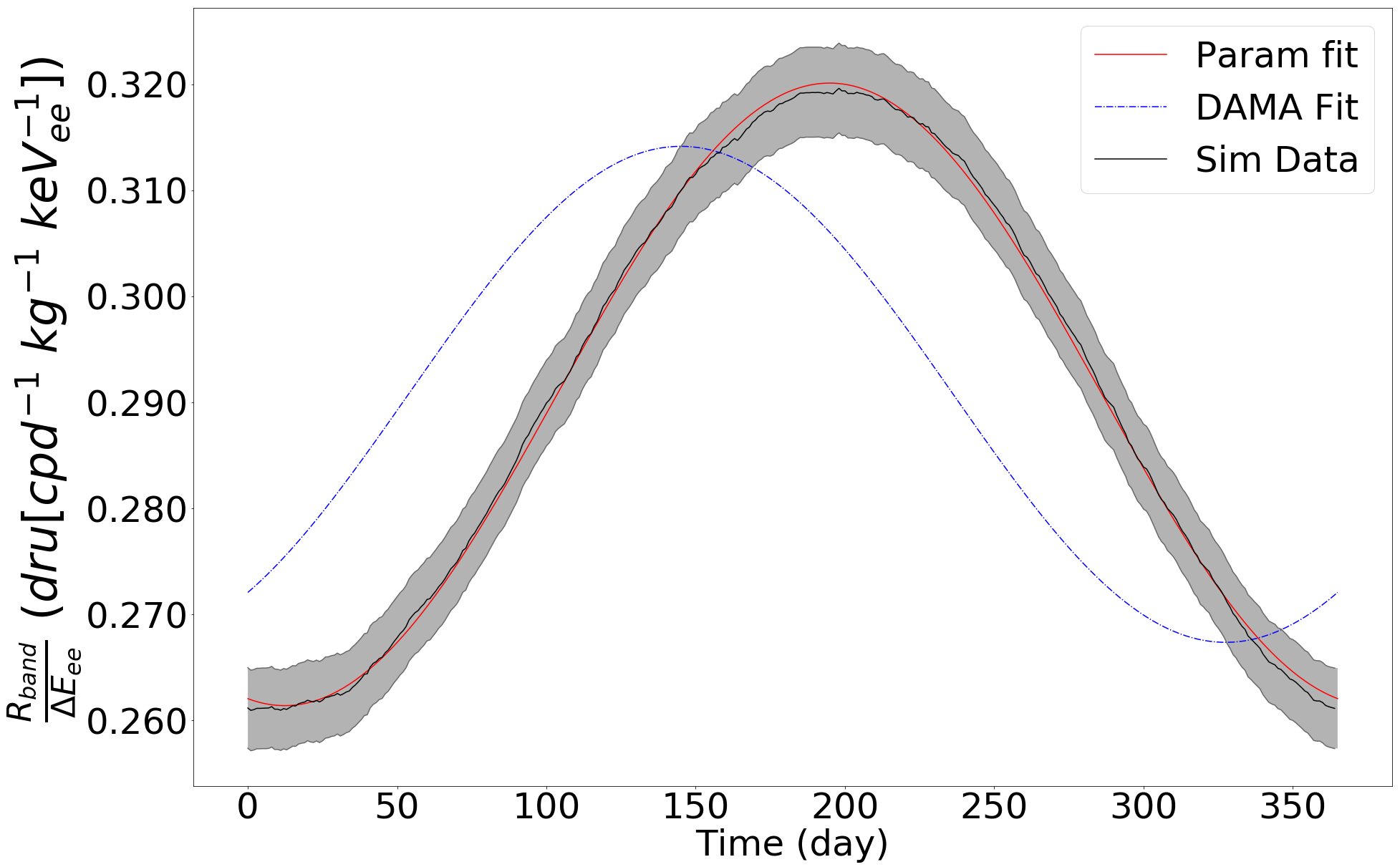}
     \caption{Annual modulation curves for the Solar Circle, evaluated per nucleon for the LMM Sodium Iodide detector between 2-6\kevee. The blue line shows a comparison to reported DAMA results for $S_m$, $t_0$, with $S_0$ scaled to match our data.The shaded region shows the 1$\sigma$ confidence interval.}
     \label{fig:am_NaI_lowmass_2-6keVee}
 \end{figure}

\begin{figure}
    \centering
    \includegraphics[width=1.1\linewidth]{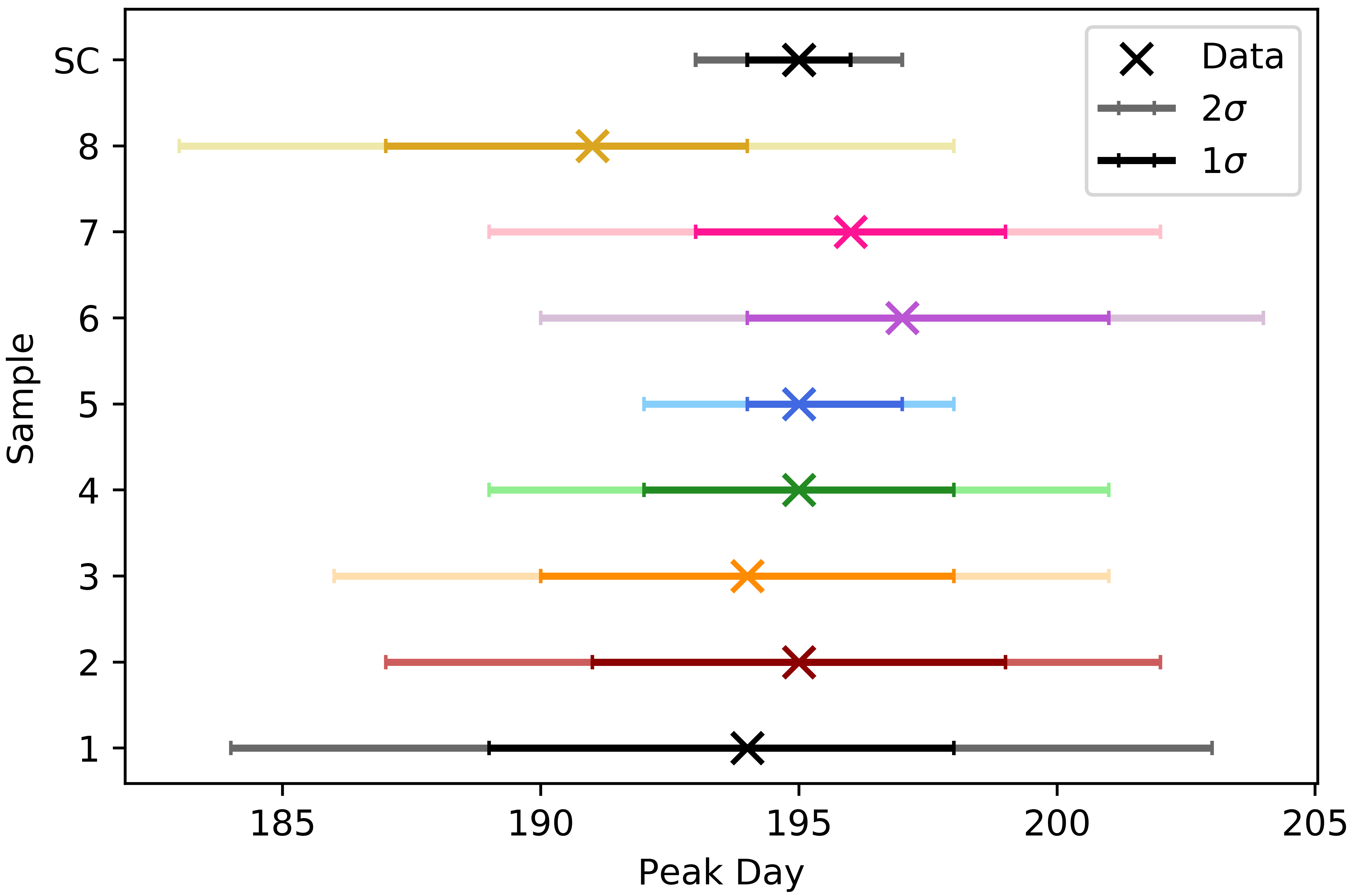}
    \caption{Peak day of dark matter counts according to the annual modulation curves of the LMM NaI detector evaluated for electron equivalent energies with 1,2$\sigma$ confidence intervals.}
    \label{fig:NaI_lowmass_peakdaye_equiv}
\end{figure}

Figure \ref{fig:am_NaI_lowmass_2-6keVee} displays the Solar Circle annual modulation predictions for the Sodium Iodide low-mass model (LMM).
The sample specific annual modulations plots can be found in \supp. The annual modulation curves demonstrate the same trends as in Section \ref{sec:Ge_LMM} and Section \ref{sec:Ge_HMM}, with fit parameters listed in Table \ref{table:NaI_fits}. The peak day plot, Figure \ref{fig:NaI_lowmass_peakdaye_equiv}, demonstrates peaks ranging from July 9-16, with 1$\sigma$ uncertainties of up to 9 days.

\subsubsection{NaI - High Mass Model}
\begin{figure}
     \centering
     \includegraphics[width=1.05\linewidth]{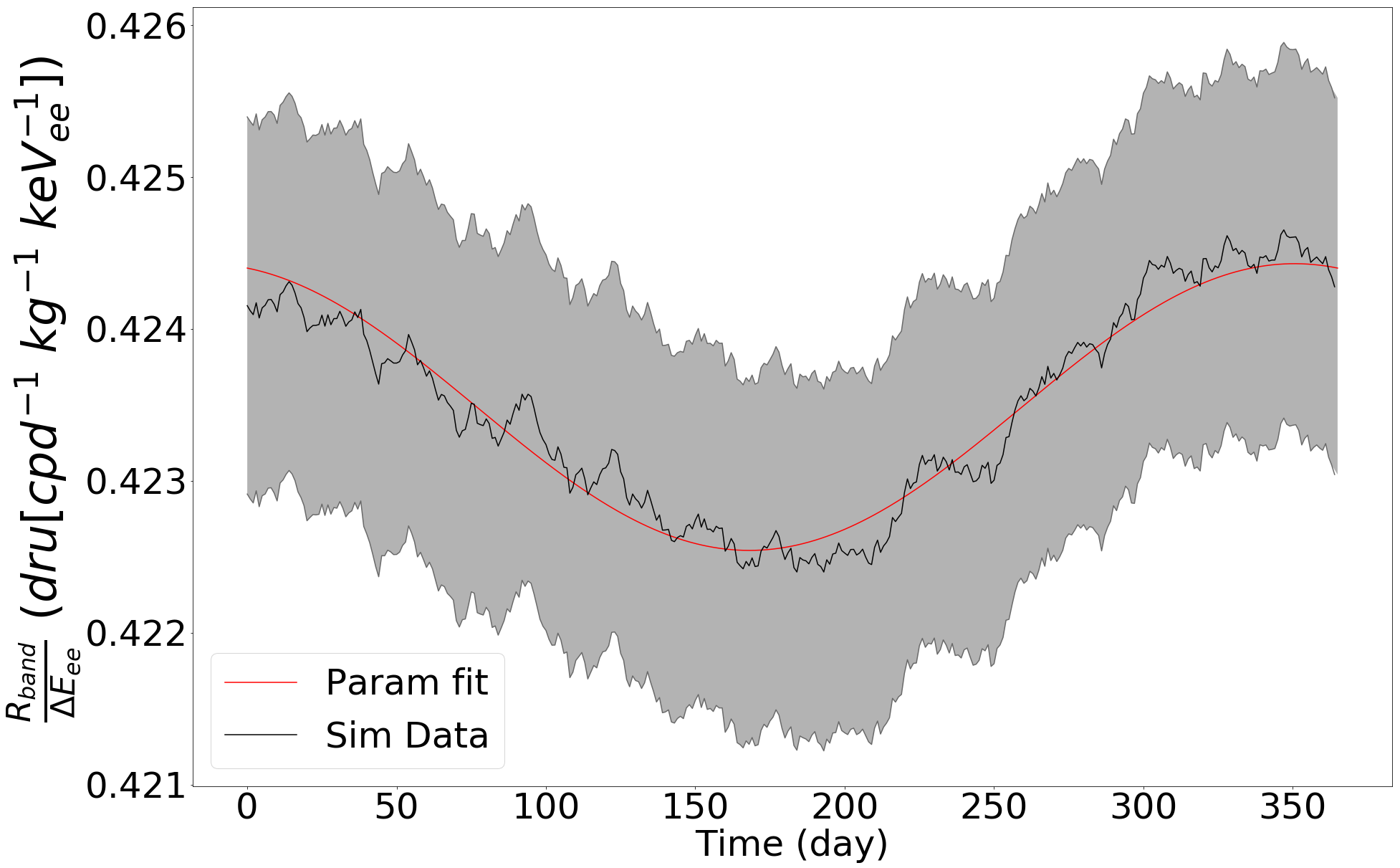}
     \caption{Annual modulation curves for the Solar Circle, evaluated per nucleon for HMM Sodium Iodide detector between 2-6\kevee with 1$\sigma$. The shaded region shows the 1$\sigma$ confidence intervals.}
     \label{fig:am_NaI_highmass_2-6keVee}
 \end{figure}

\begin{figure}
    \centering
    \includegraphics[width=1.1\linewidth]{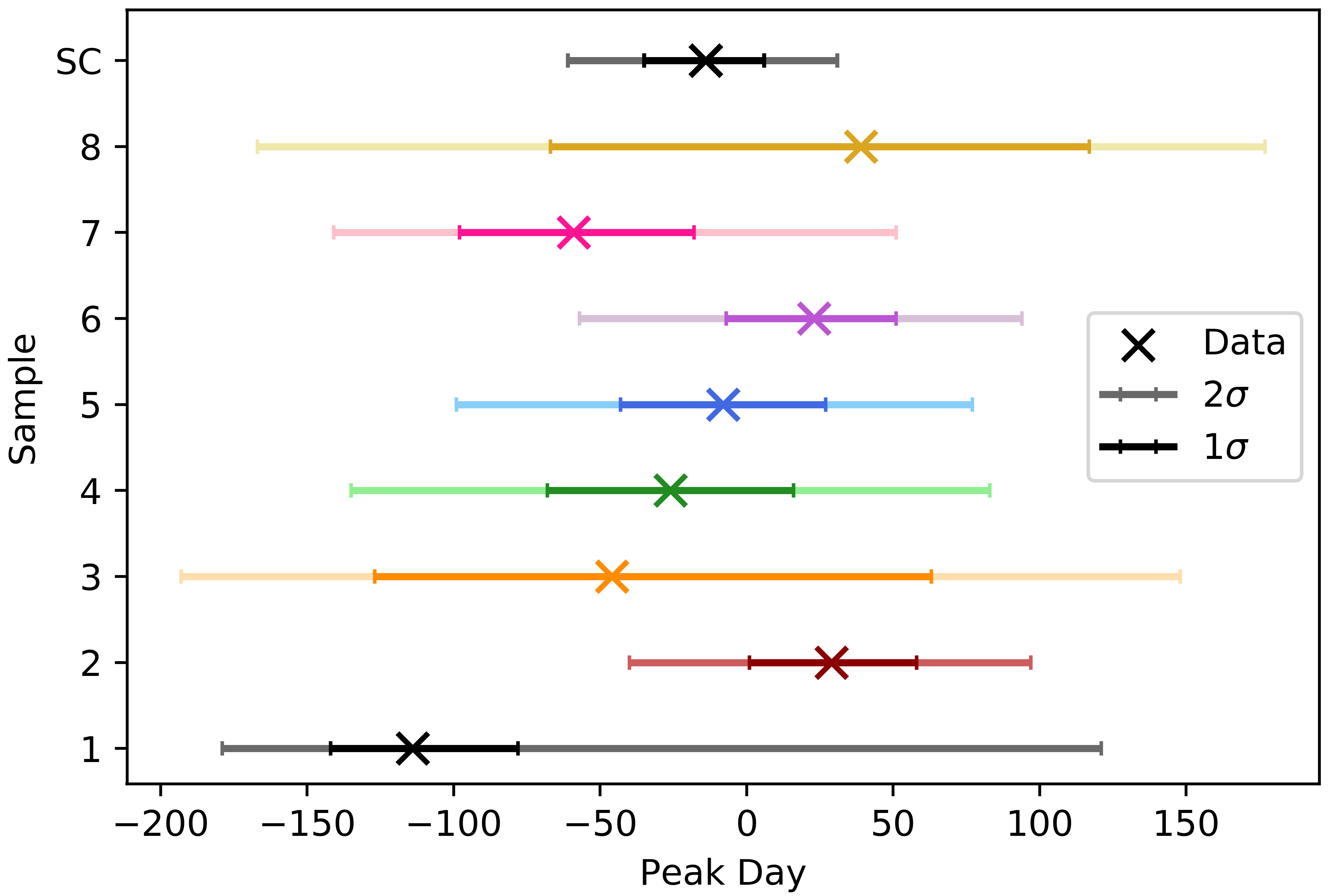}
    \caption{Peak day of dark matter counts according to the annual modulation curves of the HMM NaI detector evaluated for electron equivalent energies, with 1,2$\sigma$ errors.}
    \label{fig:Ge_highmass_peakdaye_equiv}
\end{figure}

In Figure \ref{fig:am_NaI_highmass_2-6keVee} we show the Solar Circle annual modulation predictions for the Sodium Iodide high-mass model. 
The sample specific predictions can be found in \supp. 
These demonstrate a phase flip, in the High Mass Model, as demonstrated for the Germanium results. 
For the electron equivalent energy, the 2-6\kevee region contains or is close to $Q_c$ giving the almost flat annual modulation curves evident in Figure \ref{fig:am_NaI_highmass_2-6keVee}.
Due to this, the confidence intervals in the peak day (Figure \ref{fig:Ge_highmass_peakdaye_equiv}) are enlarged with uncertainties spanning the entire year. In this scenario, the annual modulation would not be detectable.

\begin{landscape}
\begin{table}
\centering
\begin{tabular}{|l|l|l|l|l|l|l|l|l|l|} 
            & \multicolumn{3}{c}{\textbf{Nuclear Energy $E_{R}$ - 2-6\keV}} & \multicolumn{3}{c}{\textbf{Electron Equivalent Energy $E_{ee}$ - 2-6\kevee}} & \multicolumn{3}{c}{\textbf{Electron Equivalent Energy MB Fit $E_{ee}$ - 2-6\kevee}} \\
            & {\textbf{$S_0$ [\dru]}}     & {\textbf{$S_m$[mdru]}}    & {\textbf{$t_0$ [day]}} & {\textbf{$S_0$ [\dru]}}     & {\textbf{$S_m$[mdru]}}    & {\textbf{$t_0$ [day]}} & {\textbf{$S_0$ [\dru]}}     & {\textbf{$S_m$[mdru]}}    & {\textbf{$t_0$ [day]}} \\ \hline
\textbf{S1} &	$1.034_{0.019}^{0.019}$	&	$29.131_{2.985}^{3.250}$ &	$198.026_{6.049}^{6.145}$ &	$	0.291_{0.014}^{0.014}$	&	$26.226	_{1.862}^{1.978}$   &	$193.518_{4.483}^{4.485}$ &$	0.302_{0.014}^{0.015}$ & $30.190_{1.999}^{2.155} $ & $197.481_{3.759}^{3.623} $\\	\hline
\textbf{S2} &	$1.057_{0.018}^{0.017}$	&	$25.651_{2.931}^{3.227}$ &	$197.372_{6.964}^{6.743}$ &	$	0.284_{0.013}^{0.013}$	&	$27.861_{1.838}^{1.992}$    &	$194.811_{3.917}^{3.778}$ & $0.276_{0.014}^{0.014}$ & $	24.736_{1.756}^{1.825} $ & $191.908_{4.148}^{3.970}$\\	\hline
\textbf{S3} &	$1.061_{0.018}^{0.018}$	&	$20.969_{2.800}^{3.214}$ &	$196.568_{8.452}^{8.344}$ &	$0.283_{0.013}^{0.013}$	&	$27.927_{1.775}^{1.895}$    &	$193.807_{3.818}^{3.748}$ & $0.272_{0.013}^{0.013}$ & $	24.477_{1.782}^{2.078} $ & $196.684_{5.640}^{5.769} $\\	\hline
\textbf{S4} &	$1.062_{0.018}^{0.019}$	&	$22.915_{2.770}^{3.305}$ &	$193.033_{7.978}^{7.800}$ &	$0.316_{0.015}^{0.015}$	&	$33.010_{1.971}^{2.065}$    &	$195.100_{3.073}^{2.990}$ & $0.297_{0.015}^{0.015}$ & $	29.050_{1.856}^{2.035} $ & $196.045_{3.957}^{3.975} $\\	\hline
\textbf{S5} &	$1.061_{0.008}^{0.008}$	&	$25.503_{1.327}^{1.424}$ &	$197.472_{3.072}^{3.086}$ &	$	0.298_{0.006}^{0.006}$	&	$28.959_{0.816}^{0.867}$    &	$195.231_{1.640}^{1.633}$ & $0.286_{0.006}^{0.006}$ & $28.073_{0.918}^{0.953} $ & $194.106_{1.885}^{1.846}$\\	\hline
\textbf{S6} &	$1.050_{0.017}^{0.018}$	&	$30.090_{2.967}^{3.273}$ &	$200.290_{6.030}^{6.003}$ &	$	0.271_{0.013}^{0.013}$	&	$31.998	_{1.827}^{1.985}$    &	$197.199_{3.407}^{3.314}$ &  $0.287_{0.014}^{0.014}$ & $	29.592_{2.009}^{2.097} $ & $193.690_{4.583}^{4.489} $\\	\hline
\textbf{S7} &	$1.029_{0.018}^{0.018}$	&	$24.367_{2.843}^{3.382}$ &	$194.455_{7.591}^{7.583}$ &	$	0.272_{0.014}^{0.013}$	&	$32.324_{1.950}^{2.056}$    &	$196.006_{3.205}^{3.194}$ & $0.266_{0.014}^{0.014}$ & $	28.830_{1.835}^{2.026} $ & $205.355_{3.455}^{3.407} $\\	\hline
\textbf{S8} &	$1.030_{0.018}^{0.018}$	&	$24.919_{2.950}^{3.414}$ &	$195.126_{6.979}^{7.151}$ &	$	0.279_{0.014}^{0.014}$	&	$28.328_{1.797}^{1.953}$    &	$190.578_{3.918}^{3.788}$ &$	0.253_{0.013}^{0.013}$ & $	27.525_{1.899}^{2.000} $ & $196.308_{3.694}^{3.522}$\\	\hline
\textbf{SC} &	$1.052_{0.005}^{0.005}$	&	$25.446_{0.866}^{0.892}$  &	$196.979_{1.945}^{2.056}$ & $0.291_{0.004}^{0.004}$	&	$29.368_{0.532}^{0.561}$  &	$194.837_{1.046}^{1.063}$ &	$0.283_{0.004}^{0.004}$ &	$28.241_{0.600}^{0.606}$ &	$195.477_{1.208}^{1.183}$ \\\hline 
\end{tabular}
\end{table}

\begin{table}
\centering
\begin{tabular}{|l|l|l|l|l|l|l|l|l|l|} 
            & \multicolumn{3}{c}{\textbf{Nuclear Energy $E_{R}$ - 2-6\keV}} & \multicolumn{3}{c}{\textbf{Electron Equivalent Energy $E_{ee}$ - 2-6\kevee}} & \multicolumn{3}{c}{\textbf{Electron Equivalent Energy MB Fit $E_{ee}$ - 2-6\kevee}}\\
            & {\textbf{$S_0$ [\dru]}}     & {\textbf{$S_m$[mdru]}}    & {\textbf{$t_0$ [day]}} & {\textbf{$S_0$ [\dru]}}     & {\textbf{$S_m$[mdru]}}    & {\textbf{$t_0$ [day]}} & {\textbf{$S_0$ [\dru]}}     & {\textbf{$S_m$[mdru]}}    & {\textbf{$t_0$ [day]}} \\ \hline
\textbf{S1} &	$0.247_{0.003}^{0.003}$	&	$5.817_{0.740}^{0.843}$ &	$17.878_{6.820}^{9.508}$ &	$0.421_{0.004}^{0.004}$	&	$2.221_{0.820}^{1.472}$ &	$251.252_{30.993}^{31.990}$ & $0.424_{0.004}^{0.004}$ & $	1.949_{0.665}^{1.441}$ & $	19.806_{12.714}^{299.810} $\\	\hline
\textbf{S2} &	$0.243_{0.003}^{0.003}$	&	$4.020_{0.698}^{0.808}$ &	$15.793_{4.323}^{349.207}$ &	$0.423_{0.004}^{0.004}$	&	$2.342_{0.773}^{1.308}$ &	$29.412_{15.037}^{68.669}$ &$0.415_{0.004}^{0.004}$ & $	1.589_{0.495}^{1.564} $ & $	47.119_{14.441}^{281.481} $\\	\hline
\textbf{S3} &	$0.244_{0.003}^{0.003}$	&	$3.496_{0.687}^{0.823}$ &	$365.000_{352.089}^{0.000}$ &	$0.424_{0.004}^{0.004}$	&	$0.544_{0.167}^{1.810}$	&	$319.482_{260.469}^{11.222}$ &$0.423_{0.004}^{0.004}$ & $	3.356_{0.952}^{1.307} $ & $	338.609_{338.609}^{7.438} $\\	\hline
\textbf{S4} &	$0.242_{0.003}^{0.003}$	&	$3.312_{0.683}^{0.844}$ &	$25.701_{11.930}^{13.745}$ &	$0.422_{0.004}^{0.004}$	&	$1.470_{0.435}^{1.519}$ &	$0.000_{2.588}^{337.565}$ &
$0.416_{0.004}^{0.004}$ & $3.136_{0.899}^{1.370} $ & $	354.445_{351.941}^{1.701} $\\	\hline
\textbf{S5} &	$0.243_{0.001}^{0.001}$	&	$4.033_{0.329}^{0.361}$ &	$18.832_{4.676}^{4.901}$ &	
$0.422_{0.002}^{0.002}$	&	$0.874_{0.299}^{0.607}$ &	$357.376_{320.548}^{6.065}$ &
$0.423_{0.002}^{0.002}$ & $1.926_{0.433}^{0.561}$ & $351.341_{351.341}^{5.204} $\\	\hline
\textbf{S6} &	$0.247_{0.003}^{0.003}$	&	$5.817_{0.684}^{0.782}$ &	$17.244_{6.303}^{7.890}$ &	
$0.429_{0.004}^{0.004}$	&	$2.084_{0.688}^{1.499}$ &	$365.000_{334.503}^{0.000}$ &$0.423_{0.004}^{0.004}$ & $	2.509_{0.783}^{1.375} $ & $	342.093_{54.618}^{18.396} $\\	\hline
\textbf{S7} &	$0.247_{0.003}^{0.003}$	&	$5.193_{0.628}^{0.759}$ &	$19.636_{8.370}^{9.017}$ &	$0.429_{0.004}^{0.004}$	&	$1.793_{0.593}^{1.404}$ &	$306.072_{49.290}^{32.723}$ &
$0.422_{0.004}^{0.004}$ & $	3.279_{0.886}^{1.317}$ & $8.839_{2.168}^{353.835}$\\	\hline
\textbf{S8} &	$0.248_{0.003}^{0.003}$	&	$4.346_{0.772}^{0.947}$ &	$365.000_{354.232}^{0.000}$ &	$0.424_{0.004}^{0.004}$	&	$0.477_{0.229}^{1.899}$ &	$365.000_{327.905}^{30.208}$ &
$0.421_{0.004}^{0.004}$ & $	1.570_{0.499}^{1.456}$ & $18.626_{11.608}^{290.815}$\\	\hline
\textbf{SC} &	$0.244_{0.001}^{0.001}$	&	$4.332_{0.219}^{0.227}$  &	$16.885_{2.912}^{2.899}$ & $0.423_{0.001}^{0.001}$	&	$0.944_{0.263}^{0.373}$  &	$350.578_{31.460}^{14.422}$ &	

$0.422_{0.001}^{0.001}$ &	$1.273_{0.294}^{0.365}$ &	$352.872_{40.682}^{9.949}$ \\\hline
\end{tabular}
\caption{Parameter fits from Equation \ref{eq:DAMA_fit} with 1 $\sigma$ errors quoted. Classified for the LMM (top) and HMM (bottom) for a Sodium Iodide detector, and evaluated for (left) nuclear recoil energies, (centre) electron equivalent energies and (right) electron equivalent energies for Maxwell Boltzmann fits. $S_0$ values are expressed in dru units, $S_m $ in milli-dru units and $t_0$ in days.}\label{table:NaI_fits}
\end{table}
\end{landscape}

\subsection{Maxwell Boltzmann Comparison}
In order to compare the results from realistic galaxy simulations with what is predicted by the SHM, we repeated our analysis substituting $f(v,v_E)$ in Equation \ref{eq:rate_sum} with the Maxwell Boltzmann VDFs fit to the simulation velocity distribution. After fitting the simulation's galactocentric VDF with a normal distribution, then re-populating a Gaussian distribution with the fits, these new VDFs were used to calculate the expected event rates for the LMM and HMM Germanium and Sodium Iodide detectors. The simulation velocity distributions are fit to a Gaussian in each of the vector components, $v_x$, $v_y$, $v_z$, listed in Table \ref{tab:MB_v0_fits}. 

\begin{figure}
    \centering
    \includegraphics[width=\linewidth]{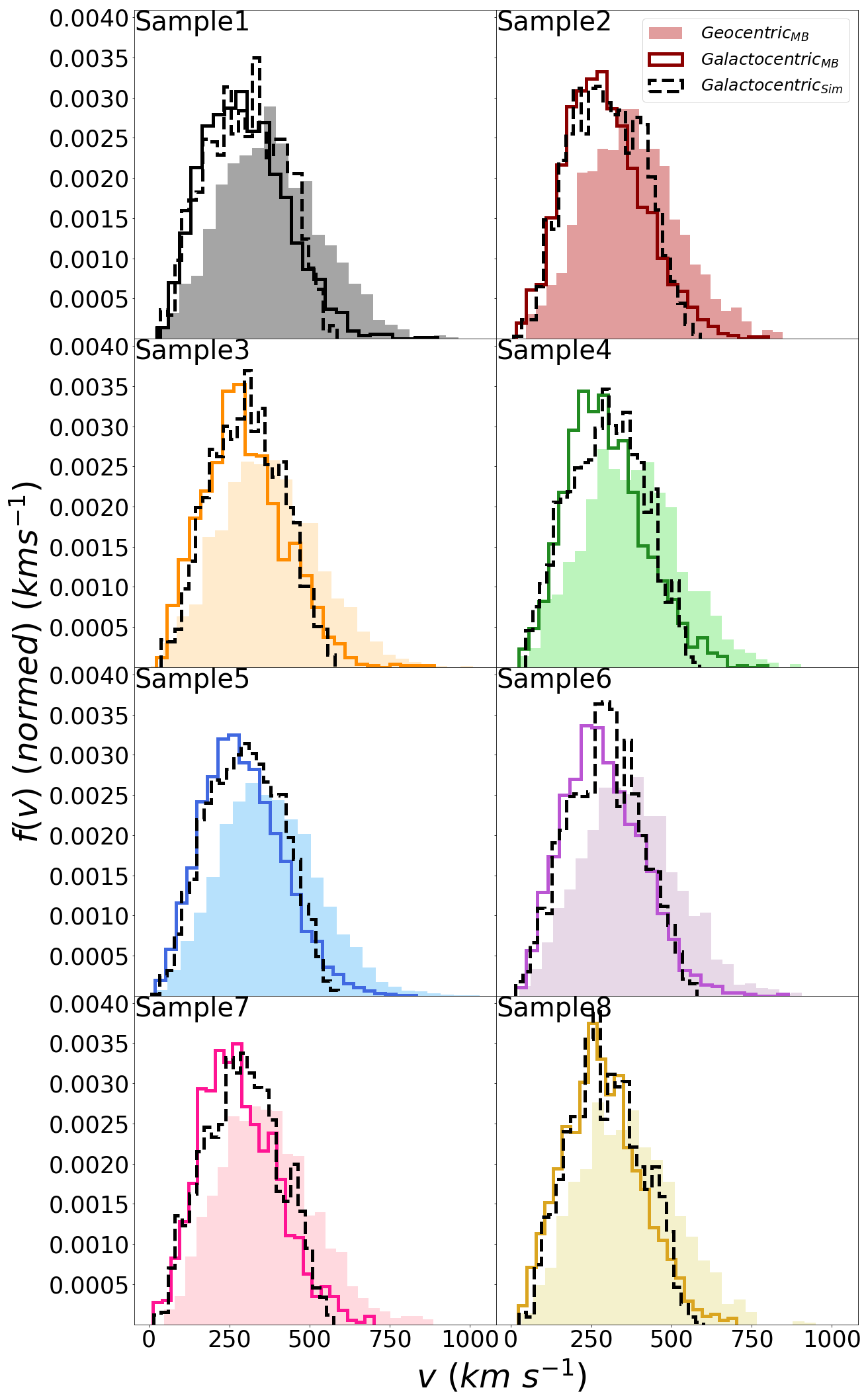}
    \caption{The velocity distribution functions for each sample around the Solar Circle. Dashed lines represent the simulation's VDF in the galactocentric frame (as per Figure \ref{fig:VDF_All_Samples} which also demonstrates a Maxwell Boltzmann fit). The dashed line represents the VDF generated by populating a Gaussian distribution using fit values from fitting a Maxwell Boltzmann function to the simulation values. The geocentric distribution is the subsequent velocities once the Maxwellian fit distribution undergoes the frame transformation into the lab frame.}
    \label{fig:MB_VDF_Comp}
\end{figure}
The velocity distributions exhibited in Figure \ref{fig:MB_VDF_Comp} demonstrate very close agreement between the galactocentric velocities in the simulation, and a Maxwell Boltzmann populated by best-fits to the simulation arrays. The geocentric distributions in Figure \ref{fig:MB_VDF_Comp} demonstrate the lab frame velocities resulting from the Maxwellian fit which are visibly smoothed with substructure effects reduced compared to Figure \ref{fig:VDF_All_Samples}.

\begin{table}
\begin{center}

\begin{tabular}{|l|lll}
\hline
\textbf{Sample} & \multicolumn{1}{l|}{\textbf{$\sigma_{v_x}$}} & \multicolumn{1}{l|}{\textbf{$\sigma_{v_y}$}} & \multicolumn{1}{l|}{\textbf{$\sigma_{v_z}$}} \\ \hline
\textbf{1}  & 202.37 & 182.40 & 171.16 \\ 
\textbf{2}  & 195.56 & 179.77 & 173.11 \\ 
\textbf{3}  & 196.89 & 184.78 & 176.95 \\ 
\textbf{4}  & 191.16 & 189.61 & 170.62 \\ 
\textbf{5}  & 196.43  & 184.78 & 172.85  \\ 
\textbf{6}  & 193.94 & 175.88 & 169.05  \\ 
\textbf{7}  & 189.10 & 180.96 & 170.92  \\ 
\textbf{8}  & 191.22 & 180.37 & 173.64 \\ 
\end{tabular}
\caption{Gaussian distribution fits used in the Maxwell Boltzmann fit comparison, given in \kms. These fits all assume a mean of $\mu = 0$.}\label{tab:MB_v0_fits}
\end{center}
\end{table}

\begin{figure*}
    \centering
    \includegraphics[width=1\linewidth]{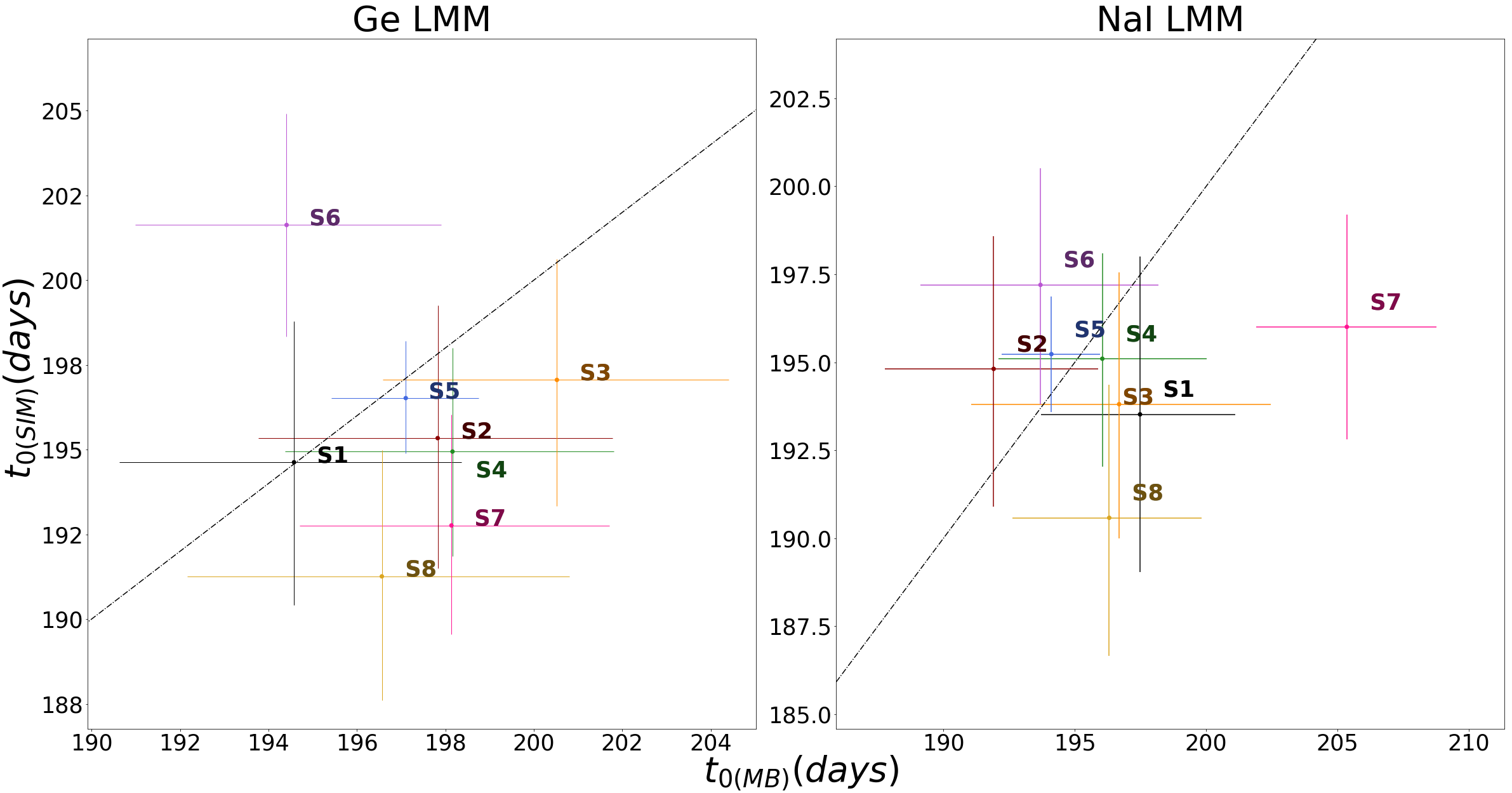}
     \caption{Peak day of dark matter counts for Ge (NaI) Low Mass DM model in the left (right) panel; according to the annual modulation curves for different samples against the parameter fit when using the Maxwell Boltzmann realisation for those VDFs. The one to one peak day is given by the dotted line.}\label{fig:t0_comp}
\end{figure*}

Tables \ref{table:Ge_fits} and \ref{table:NaI_fits} list the $S_0$, $S_m$, $t_0$ fit values for the Maxwell Boltzmann distribution cases.  When we compare the peak day parameter $t_0$ for each sample between the simulation results and the Maxwell Boltzmann analysis results we find agreement between both Germanium and Sodium Iodide for the case of the Low Mass Model ($\ \sigma_0 = 1.3 \times 10^{-41} {\rm cm}^2$; $M_D = 15$\gevc) as shown in Figure \ref{fig:t0_comp}.
The agreement between the peak-day fits of the simulation VDFs and the Maxwell Boltzmann fits to the simulation VDFs suggest that the substructure in the `messy' halos doesn't significantly shift the peak days within the precision of our measurements.

\section{Discussion}\label{sec:Discussion}
In this paper we have determined predictions for terrestrial dark matter detectors using simulations of unprecedented resolution run with the FIRE-2 physics model \citep{10.1093/mnras/sty1690}, offering high particle resolution and an extension on previous treatments of modelling dark matter direct detection. 
We demonstrate our approach using two different detectors, Germanium and Sodium Iodide, whose analysis highlights the effects of single versus compound detectors, and provides a better understanding of the capabilities and limitations of different detector materials, for varying dark matter models. This work has the unique advantage of the FIRE-2 simulations which combine sub-resolution feedback processes to provide Milky Way-type halos with resolutions surpassing previous efforts. Our analysis technique provides the machinery to use the full particle velocity distribution to compute event rates, in order to eliminate dependencies on the SHM usually present in past works.

As outlined in previous literature \citep{PhysRevD66083003,Kuhlen_2010}, we find strong deviations of the lab-frame velocity distribution of the incoming dark matter headwind from the traditional Maxwell Boltzmann form. 
The baryonic backreaction and gravitation of the dark matter near to the disk within the hydrodynamic simulation further complicates these deviations \citep{Duffy_2010,Bryan_2013}. They present as structure in the high-velocity tail of the lab-frame velocity distribution functions, indicating the potential presence of streams and debris flows within the galaxy, as has already been confirmed by \cite{Helmi_2018,10.1093/mnras/sty982, O_Hare_2020} for the Milky Way and \cite{Sanderson_2020, 2021ApJ...920...10P} for the Latte suite, and specifically for the halo m12f we analyse in our study. 

The propagation of substantial sub-structure in the Solar Circle to minor scatter in the consequent detection rates suggests that parallel conclusions would be drawn for other halos in the Latte suite, whose spatial structure, mass and morphology closely align both with the Milky Way, and with each other.  Halo m12i may offer an exception with a strong presence of stellar streams in its Solar Circle, potentially capturing higher energy effects resulting in larger deviations to the signal predictions.

In this halo, we note that there is high-velocity structure in 7 out of the 8 samples we select around the Solar Circle (see Appendix \ref{Appendix:VDF}) and that all 8 demonstrate deviations from a standard Maxwell Boltzmann distribution. These volume-limited samples are then stacked to combine into a representative `Solar Circle' sample, akin to a mass-weighted sample, which demonstrates a smoothed version of its constituents in Figure \ref{fig:VDF_Solar_Circle}. This Solar Circle sample is still not well-described by a truncated Maxwell Boltzmann distribution. These complex zoom-in simulations offer attractive opportunities to providing experiments with more realistic, i.e. messy, dark matter substructure velocities.

We investigate the effects of these realistic velocity distributions on dark matter detection using the annual modulation predictions. These find, for the experimentally significant 2-6\kevee region, that all signals recover the expected sinusoidal curve with amplitude $S_m$ values of order $\mathcal{O}(10^{-2})$ \dru\ for the Low Mass Model ($M_D=15$ \gevc, $\sigma_0=1.3\times 10^{-41} {\rm cm^{-2}}$) and $\mathcal{O}(10^{-4}-10^{-3})$ \dru\ for the High Mass Model ($M_D=60$ \gevc, $\sigma_0=5.5\times 10^{-42} {\rm cm^{-2}}$). Generating annual modulation predictions for individual samples, we find that the best-fitting sinusoidal parameters are consistent across different samples.  Hence, the impact of varying position around the Solar Circle, and the related velocity structures, doesn't cause significant dispersion. The limited effect of the velocity structure on the derived parameters is further emphasised by the agreement between fits using the full VDFs, and fits using equivalent Maxwell Boltzmann distributions, a result in agreement with \cite{Bozorgnia_2016,Bozorgnia_2020}. This small variation is a reassuring result for experimental efforts to detect dark matter directly by taking advantage of the solar system's motion around the galactic centre. 

Due to the volume-limited nature of the samples, regions of higher density will be more tightly constrained by the increased amount of particles, as in Sample 5. But we will less frequently sit within such a volume based on this example simulation.

The predicted annual modulation time series depends sensitively on the dark matter mass and the observed energy window.  We illustrate these effects through comparison of the results using nuclear recoil and electron equivalent energies, and between the low-mass and high-mass models. In the cases of the HMM, the energy region in \kevee falls more closely to the critical energy $Q_c$ and is equivalent to a larger energy range ($ 9.6-24.9$\keV $\approx 2-6$\kevee for Germanium).  In this case, the sinusoidal annual modulation amplitude is very small.

We find minimal correlation between the fitted parameters $S_0$, $S_m$, $t_0$ in Figures \ref{fig:S0_Sm_Ge_lowmass_keV} and \ref{fig:S0_Sm_Ge_lowmass_keVee}. 
The magnitudes of these fit values highlight the importance of the correct modelling of quenching factors to determine the energy of the incoming particles after observing the recoil signature left at a particular electron equivalent energy. While previous works to accurately model the scintillation and subsequent quenching factor for a range of detector materials including Ge and NaI have been studied, we provide the unique comparison between an energy range in both \keV and \kevee, highlighting how small changes in the energy region of interest can have significant impact on the expected signal features.

Focusing on the predictions of the peak day of the annual modulation, when energies approach the critical energy $Q_c$ (defined for low energy recoils of a detector as a function of dark matter mass \citep{Lewis_2004}), we observe a phase flip. 
The phase flip is evident for both Germanium and Sodium Iodide predictions in this work and occurs in the same direction when moving between Low and High Mass Models. The critical energy, which lies within this 2-6\kevee region for these element and compound detectors, implies that the success criteria for an annual modulation signal due to dark matter (peaking at $\sim t_0= 152$) is not always indicative of a detection. By informing the community of the phase flip for Germanium and Sodium Iodide in experimentally significant energy regions, we widen the possibility for the interpretation of direct detection signals. This ensures we consider additional subtlety and nuance of the physics involved, both from astrophysical and particle physics perspectives.

Comparing these predictions to \cite{Bernabei_2018} claim, we note in Figure \ref{fig:am_NaI_lowmass_2-6keVee} that the peak day is shifted forward in time compared to the DAMA results, which quotes $t_0 = 145 \pm 5$ \citep{Bernabei_2018}. Compared to the Sodium Iodide LMM in \kevee, we find a phase shift of 50 days compared to \cite{Bernabei_2018}. In part, this difference could be attributed to gravitational focusing of dark matter by the Sun which is not taken into account in this work \citep{Lee_2014} and can cause shifts in the phase by up to 21 days. The confidence intervals quoted in our tables show that we match DAMA's best uncertainty for limits. This indicates that our simulations are perfectly tuned to match current and upcoming experimental capabilities.  The \cite{Bernabei_2018} results quote $S_m = 0.0234$ \dru\ which sits in close alignment with the numbers reported in Table \ref{table:NaI_fits}.

\section{Conclusions}\label{sec:conclusions}
Using the high resolution zoom-in galaxy FIRE-2 simulations, we explored the variance across samples around the Solar Circle in the distribution of the dark matter and the resulting implications for dark matter direct detection. 
Within individual samples, the velocity distribution functions exhibited considerable high-velocity tails. 
These VDFs in the galactocentric frame are noticeably non-Maxwellian, unlike traditional distributions frequently assumed in the literature, and persisted after undergoing the Galilean boost to the lab frame in agreement with \citet{Kuhlen_2010}. These deviations were found to be a manifestation of the inherent `messiness' of the dark matter field within the Solar Circle, as no phase-coherent streams were found to coincide with the sampled regions of the Solar Circle. However, whilst these astrophysical variations affect the detailed annual modulation time-sequence, their influence on derived parameters such as the peak day is much more limited.  We find that the best-fitting peak days are consistent between different samples, and also do not change significantly if the full VDFs are replaced by equivalent Maxwell Boltzmann distributions.

We also demonstrated that the event rate predictions were very sensitive to details such as the observed energy window, the quenching factor and the dark matter mass, with the intrinsically higher number of events sitting closer to the median of the distribution dismissing higher energy structure effects and high energy events being preferentially down-weighted by the Form Factor.

The major findings of this paper were;
\begin{enumerate}
    \item The velocity distributions in the galactocentric reference frame all demonstrate significant deviations from the fiducial Maxwell Boltzmann fit, without containing phase-coherent stellar streams, demonstrating the inherent fluctuations in the dark matter field.
    \item The consequent annual modulation signals, obtained using a novel count rate approach, demonstrated agreement between samples to within 1$\sigma$ confidence, demonstrating that high energy structure effects from the velocity distribution functions do not persist. 
    \item Parallel conclusions were achieved in paramaterizing annual modulation signals obtained using Maxwell Boltzmann fits to the realistic simulation velocities implying that, to the precision of our measurements, the Standard Halo Model is an appropriate approximation. 
\end{enumerate}

For this work, we developed an analysis pipeline, \DarkMaRK, which can be used for modelling future dark matter detection experiments, for a range of dark matter models, interaction schemes and astrophysical effects.
\section*{Acknowledgements}
The author G.E.L. gratefully acknowledges Darren Croton for his guidance and feedback on our manuscript and Ciaran O'Hare for valuable scientific discussions and insights during this work.
G.E.L. would like to acknowledge support through an Australian Government Research Training Program Scholarship. This research was supported by the Australian Research Council Centre of Excellence for Dark Matter Particle Physics (CDM; project number CE200100008). Parts of this research were supported by the Australian Research Council Centre of Excellence for All Sky Astrophysics in 3 Dimensions (ASTRO 3D; project number CE170100013). Figures in this work were created using Matplotlib \citep{Hunter:2007} and made use of colormaps from the CMasher package \citep{2020JOSS....5.2004V}.

\section*{Data Availability}
The author acknowledges the use of the FIRE-2 simulations, specifically the Latte suite. The full simulation halo snapshot for the m12f halo used in this work is publicly available via yt Hub at ananke.hub.yt.
The Dark Mark python package created for this work is publicly available at https://github.com/Grace-Lawrence/Dark-MaRK.
The secondary data underlying this article will be shared on reasonable request to the corresponding author.


\bibliographystyle{mnras}
\bibliography{Gusts_in_the_Headwind} 
\clearpage

\appendix
\section{Velocity Distribution Functions for All Samples}\label{Appendix:VDF}
\begin{figure}
    \centering
    \includegraphics[width = 1.0\linewidth]{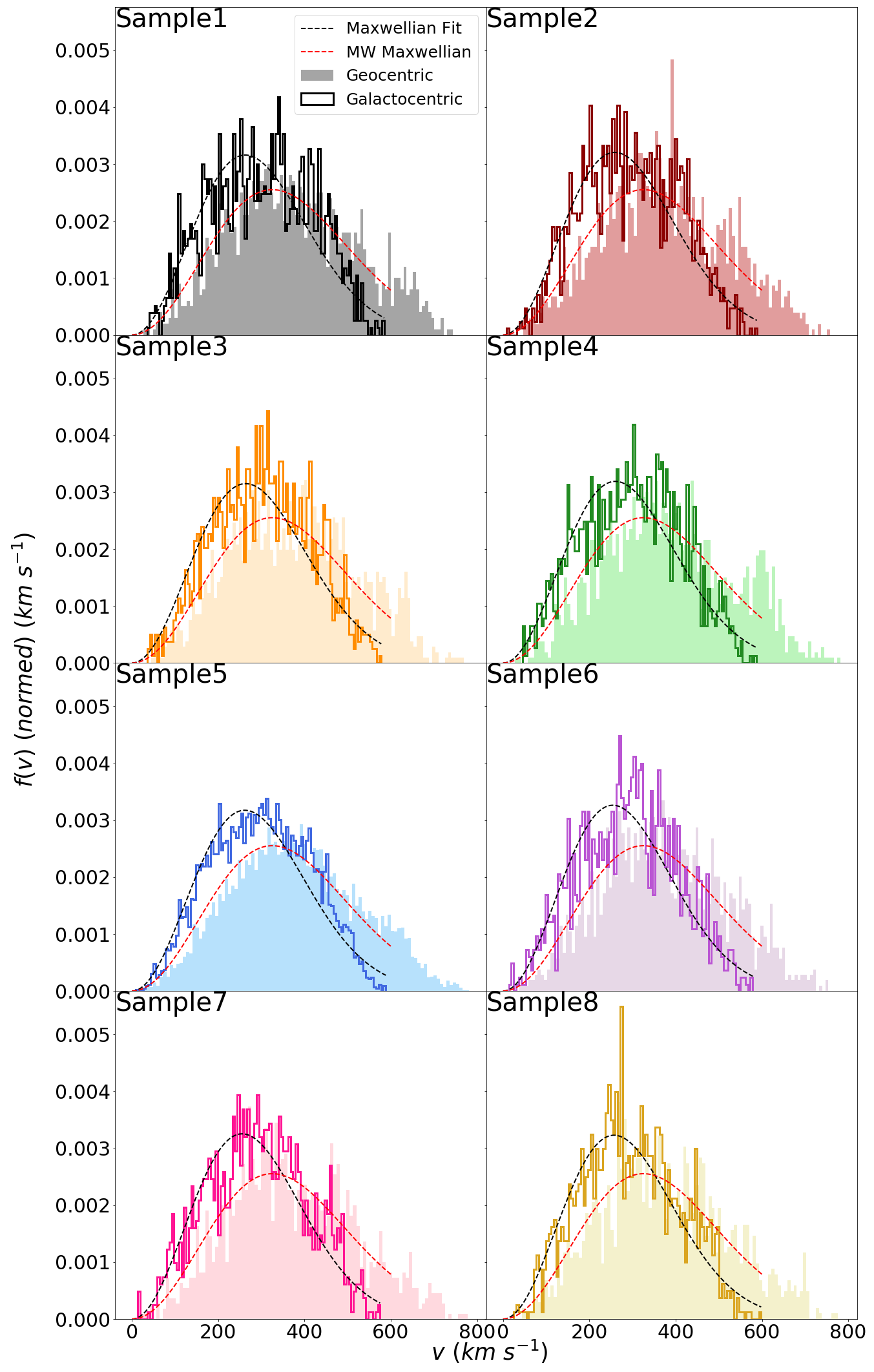}
    \caption{Galactocentric and geocentric velocity distribution functions for all samples around the Solar Circle, where the geocentric distribution is evaluated at its peak day. The black dashed line represents a Maxwell Boltzmann function fit to the galactocentric distribution. The red dashed line represents a Maxwell Boltzmann function with the standard values for the Milky Way, $\sigma= 230kms^{-1}$, $\bar{v} = 0kms^{-1}$, truncated at $600kms^{-1}$, demonstrating the deviations of our simulation velocities from fiducial assumptions.}
    \label{fig:VDF_All_Samples}
\end{figure}
Figure \ref{fig:VDF_All_Samples} demonstrates the velocity distribution functions for all eight samples about the Solar Circle, in both the galactocentric and geocentric reference frames. The geocentric frame is formed from applying a Galilean boost to the galactocentric distribution, resulting in high-velocity substructure becoming more prominent in this lab-frame.

The dashed line represents a Maxwell Boltzmann function fit to the galactocentric reference frame. There are clear deviations in the tail due to high velocity structure in the samples. However, at the highest velocity end of the tail, our simulations consistently under-predict compared to the Maxwell Boltzmann, most likely due to the escape velocity of the simulated galaxies.

\begin{figure}
    \centering
    \includegraphics[width = 1.05\linewidth]{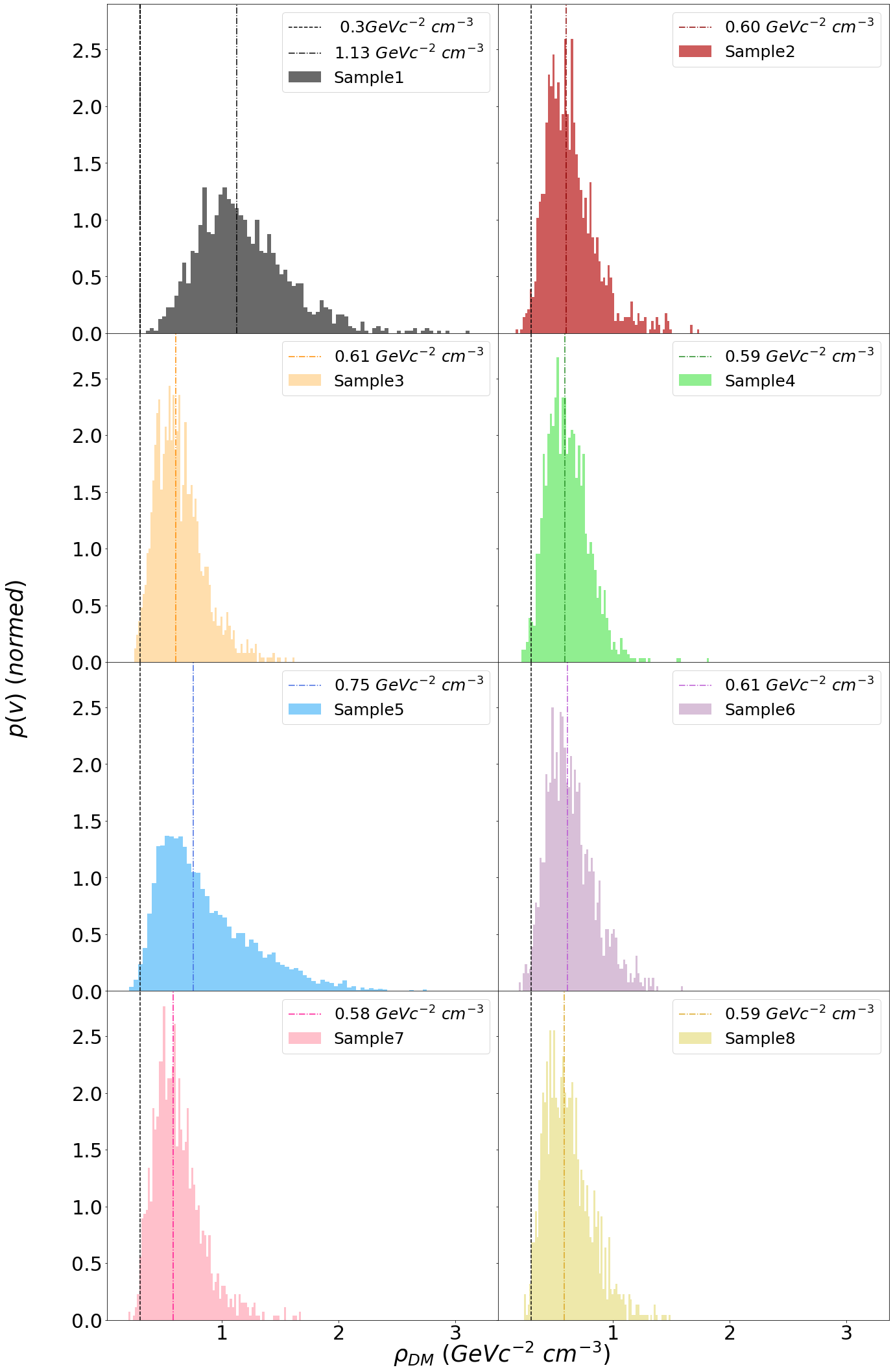}
    \caption{Normalized density histograms for each of the samples around the Solar Circle region demonstrating the distribution of densities of simulation particles within the volume. For each sample the median density value is plotted (dash-dot, colored) in addition to the standard fiducial value of 0.3\gevccm (dashed, black).}
    \label{fig:Rho_All_Samples}
\end{figure}

The variation of the velocity distribution around the Solar Circle is indicative of the velocity substructure present within the solar neighbourhood which influences expected detection rates for terrestrial dark matter searches. 
Figure \ref{fig:Rho_All_Samples} shows density histograms for the density of each of the Solar Circle samples.
Figure \ref{fig:Rho_All_Samples} demonstrates the spread of particle densities for each sample, with Sample 1 and 5 showing broader distributions and higher median density values.

\section{Nuclear Recoil Energy Evaluations}\label{app:E_NR}
Nuclear recoil energy predictions of the annual modulation signal for the Low Mass Model for Sodium Iodide detectors, and the High Mass Model for both Germanium and Sodium Iodide detectors are provided here for comparison.

\subsection{Germanium}
\subsubsection{High Mass Model}
\begin{figure}
    \centering
    \includegraphics[width=96mm]{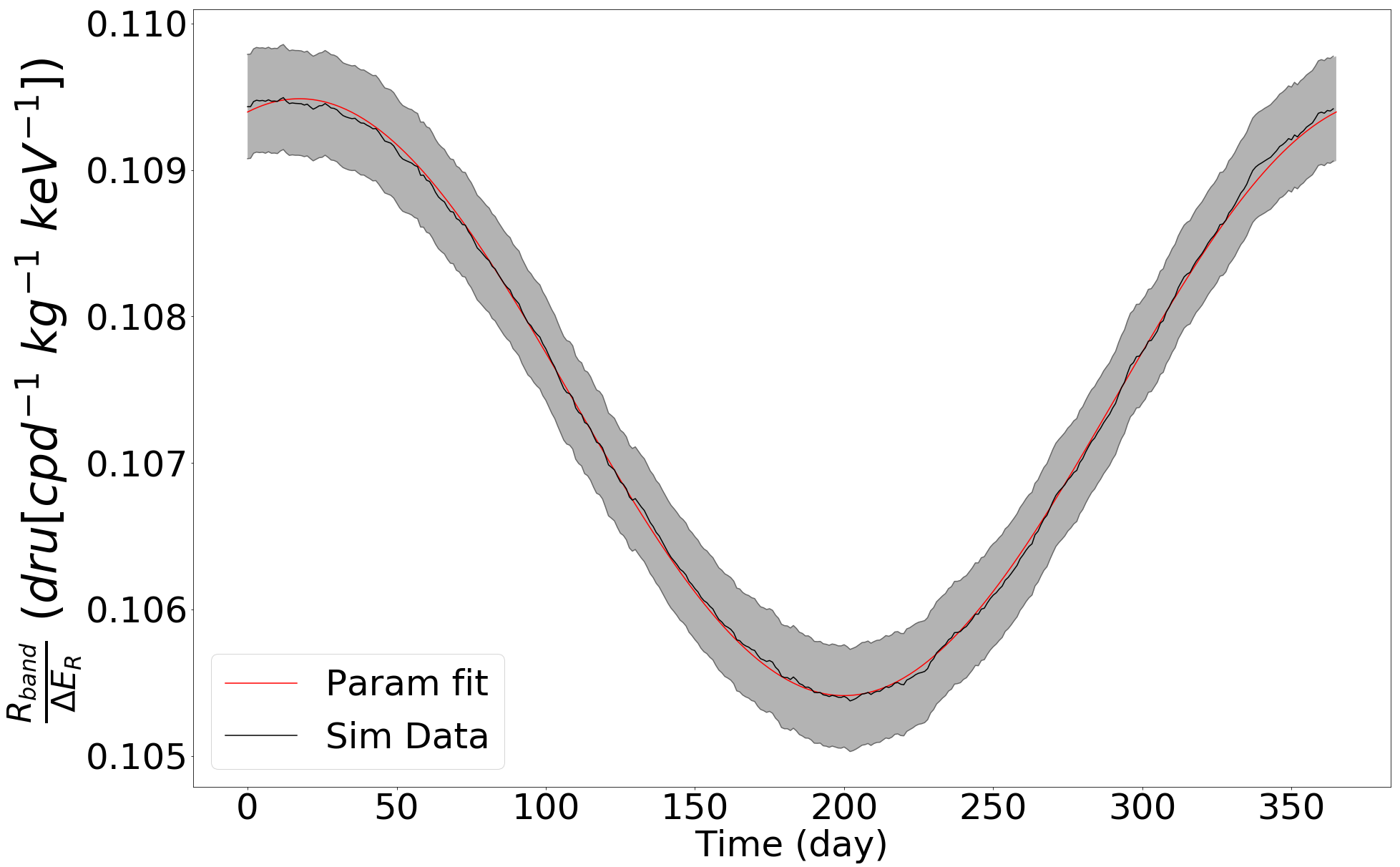}
    \caption{Annual modulation curve for the Solar Circle, evaluated per nucleon for the High Mass Model dark matter particle and a  Germanium detector. Evaluated between 2-6\keV with 1$\sigma$ CI, the red line demonstrates a fit to Equation \ref{eq:DAMA_fit}.}
    \label{fig:am_Ge_highmass_2-6_tot}
\end{figure}

\begin{figure}
\centering
\includegraphics[width=97mm]{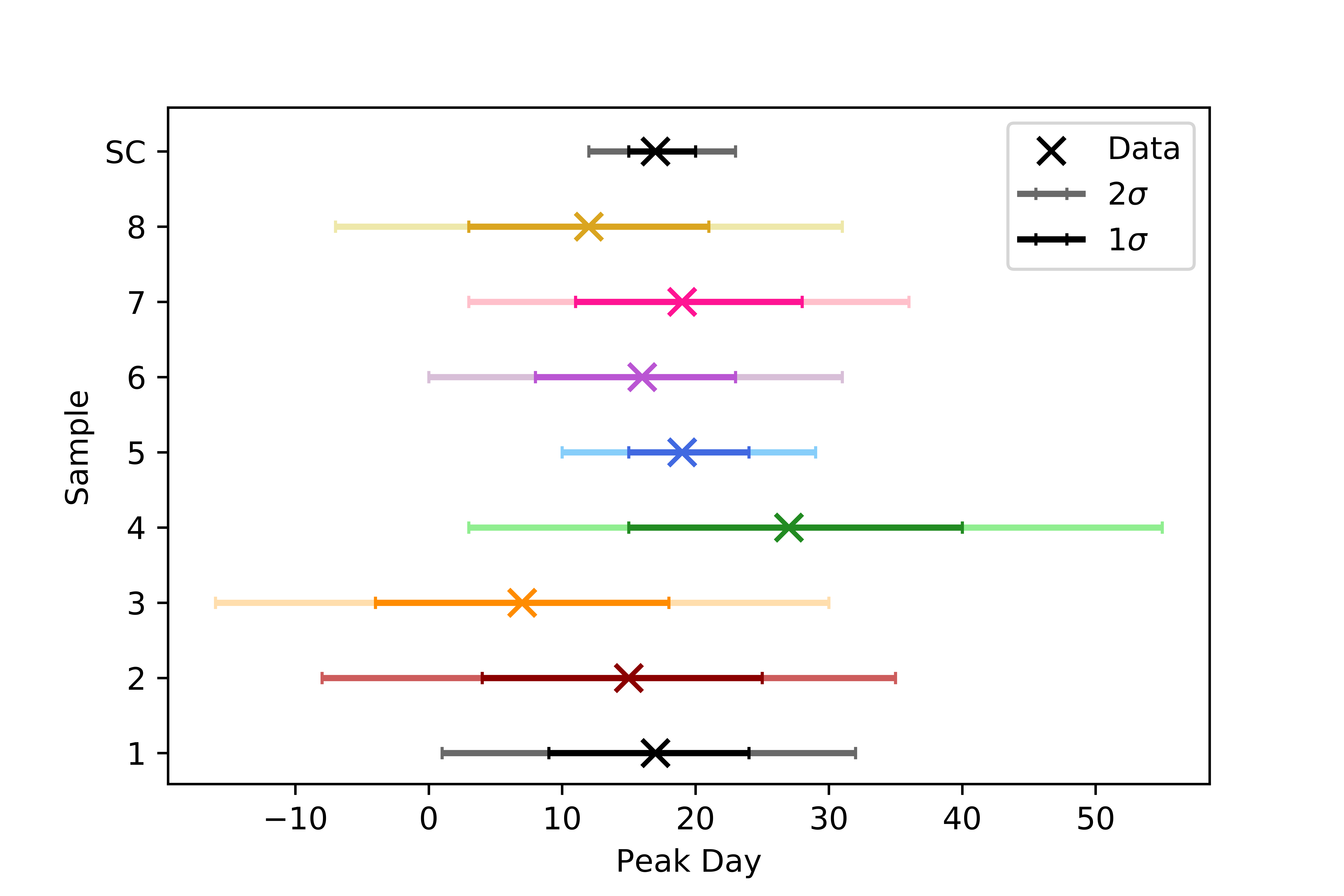}
\caption{The peak day of dark matter counts according to the annual modulation curves of the High Mass Model dark matter particle interacting with a Germanium detector. Evaluated for nuclear recoil energies with 1,2$\sigma$ errors.}\label{fig:Ge_highmass_peakday}
\end{figure}

Figure \ref{fig:am_Ge_highmass_2-6_tot} shows the annual modulation prediction for the Solar Circle sample. It has undergone the expected phase transition and peaks at the end of the year. Figure \ref{fig:Ge_highmass_peakday} shows the peak day expectations for annual modulation signals evaluated using nuclear recoil energies. The peak days span from day 365(0) to day 27.
\clearpage

\subsection{Sodium Iodide}
\subsubsection{Low Mass Model}
 \begin{figure}
     \centering
     \includegraphics[width=90 mm]{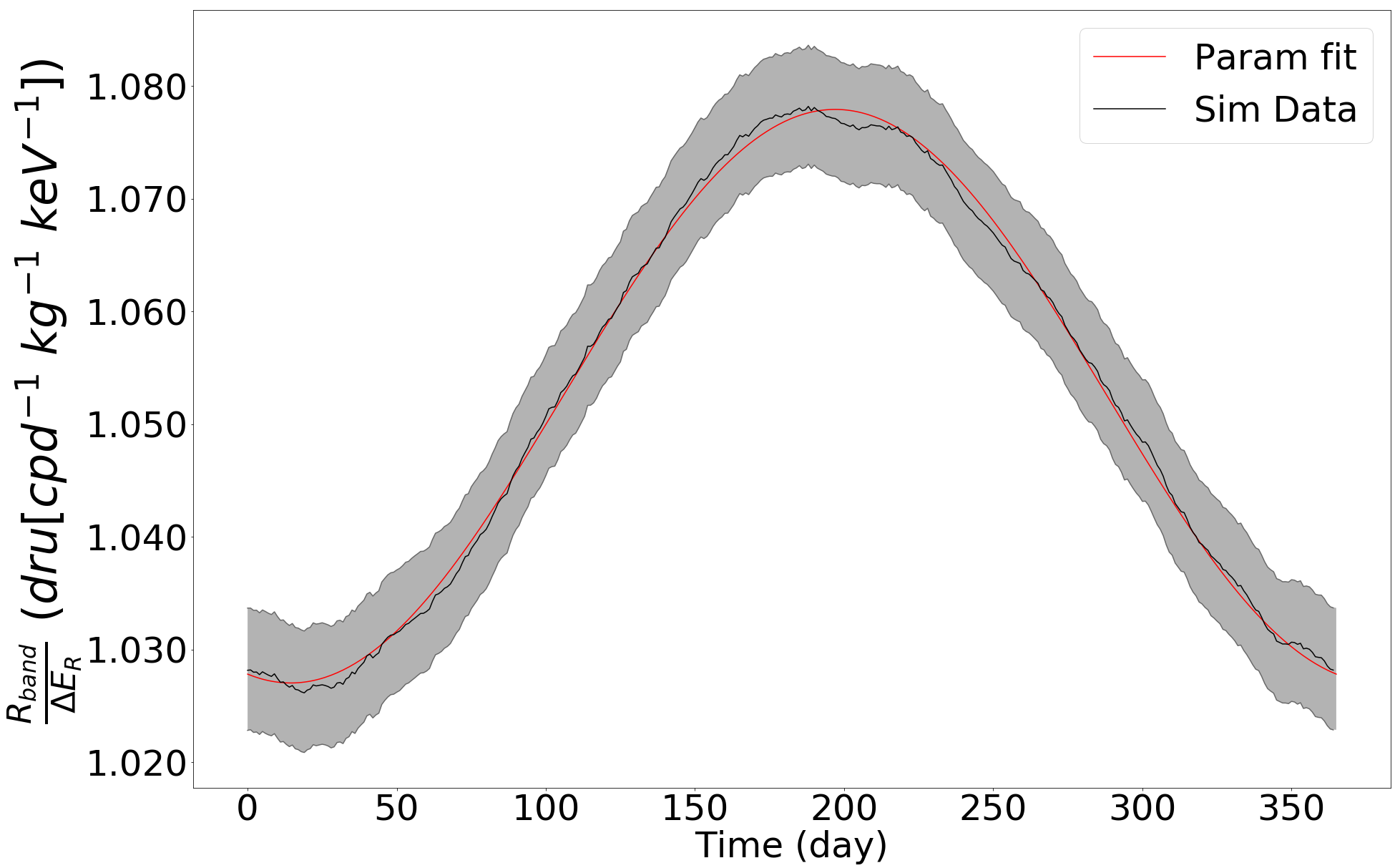}
     \caption{Annual modulation curve for the Solar Circle, evaluated per nucleon for the Low Mass Model dark matter particle interacting with Sodium Iodide between 2-6\keV with 1$\sigma$ CI.}\label{fig:NaI_LMM_SC_am}
 \end{figure}
 
 \begin{figure}
\centering
\includegraphics[width=80mm]{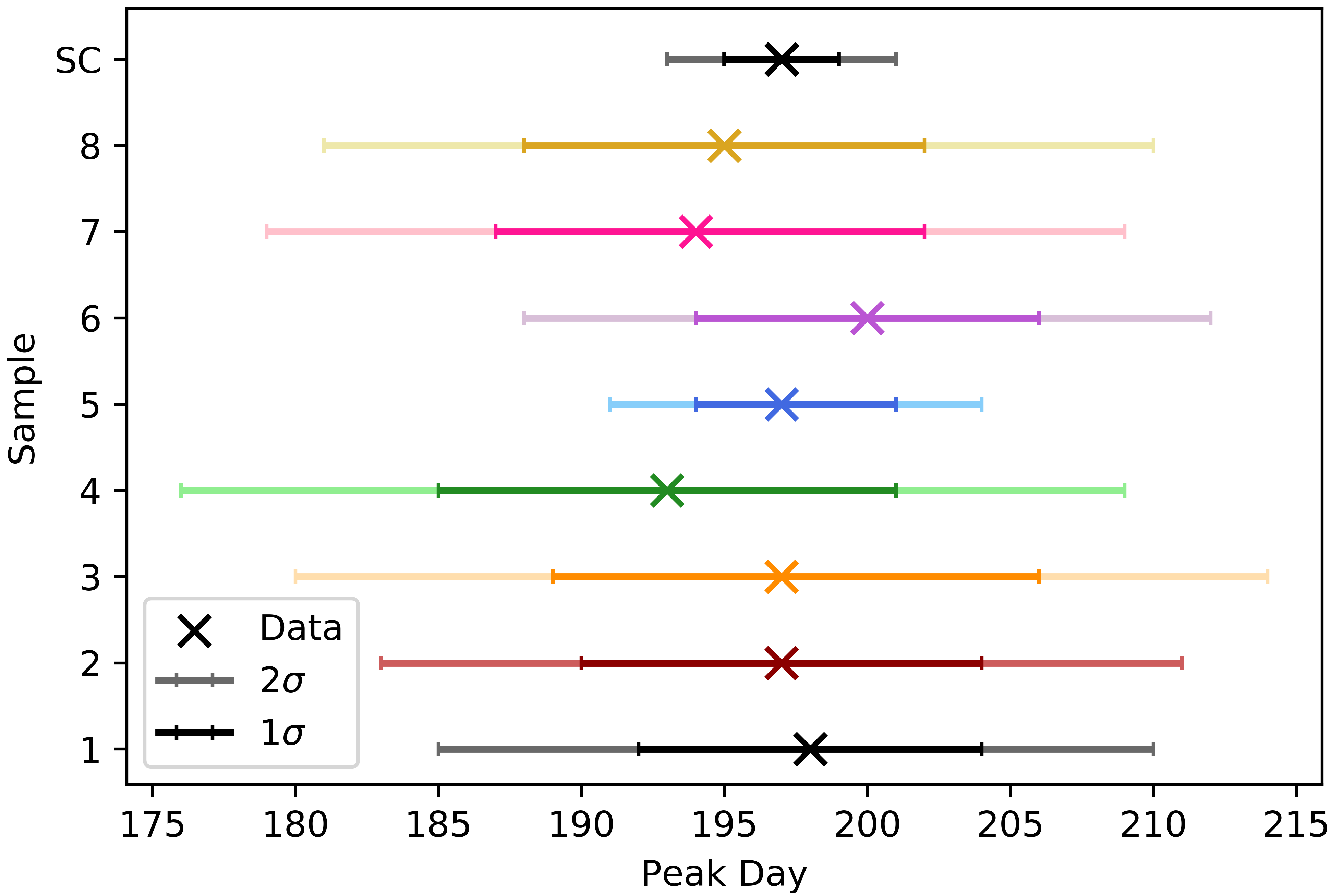}
\caption{The peak day of dark matter counts according to the annual modulation curves of the Low Mass Model dark matter particle interacting with a Sodium Iodide detector. Evaluated for nuclear recoil energies with 1,2$\sigma$ errors.}
\label{fig:NaI_lowmass_peakday_nuclear}
\end{figure}

Figure \ref{fig:NaI_LMM_SC_am} demonstrates the annual modulation curve for a Sodium Iodide detector interacting with the Low Mass Model dark matter particle. The result demonstrates a decrease in modulation amplitude of a factor $\sim$ 1.15, compared to the corresponding electron equivalent energy range.

The peak day evaluations in Figure \ref{fig:NaI_lowmass_peakday_nuclear} demonstrate the modulation peaking in the middle of the year, inline with theoretical predictions. 
Agreement between samples to within 1$\sigma$ further confirm the conclusion that the variation of dark matter about the Solar Circle has minimal impact on key signal parameters.

 \subsubsection{High Mass Model}
 \begin{figure}
     \centering
     \includegraphics[width=92mm]{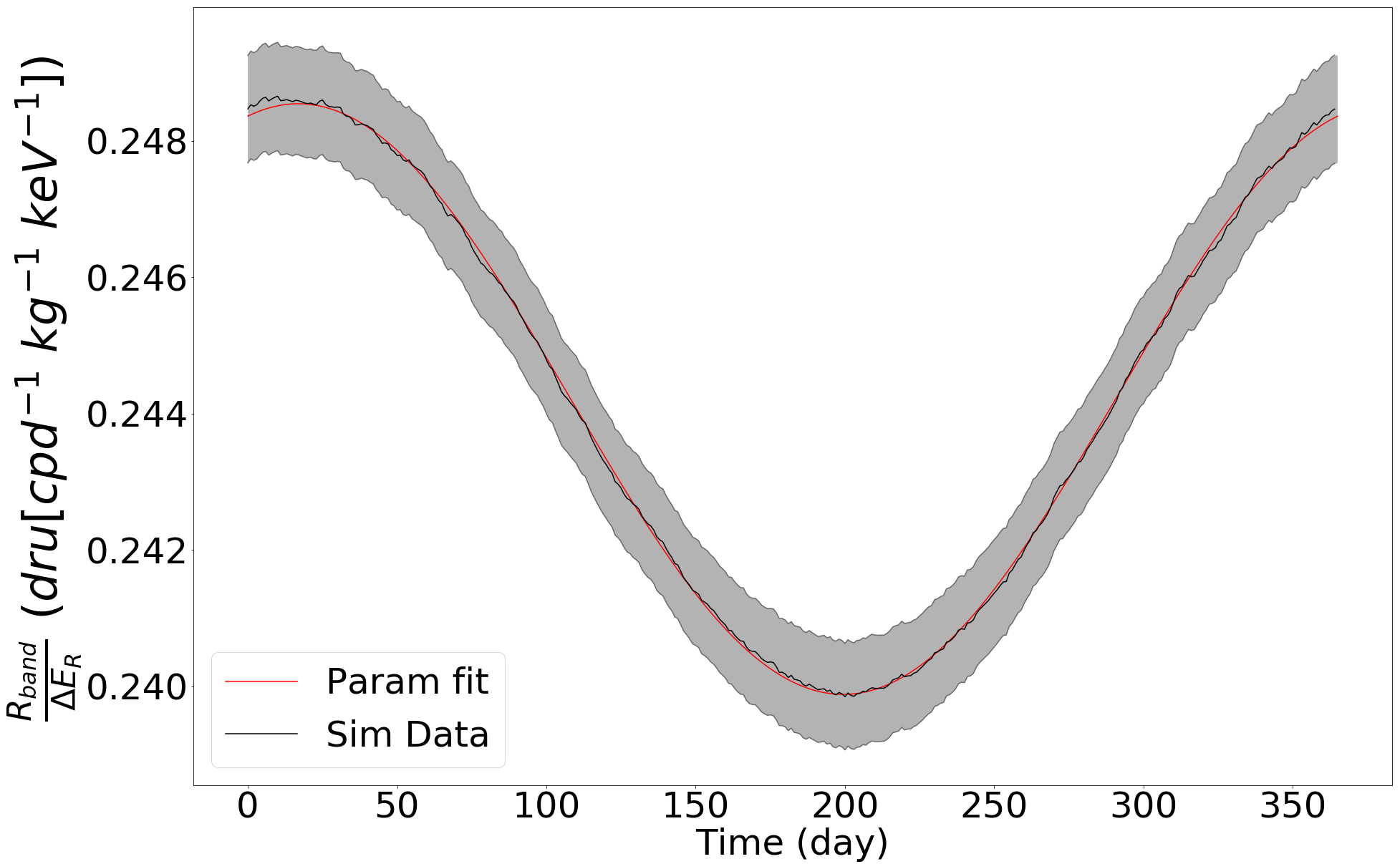}
     \caption{Annual modulation curve for the Solar Circle, evaluated per nucleon for the High Mass Model dark matter particle interacting with Sodium Iodide between 2-6\keV with 1$\sigma$ CI.}
     \label{fig:am_NaI_highmass_2-6}
 \end{figure}
 
\begin{figure}
\centering
\includegraphics[width=80mm]{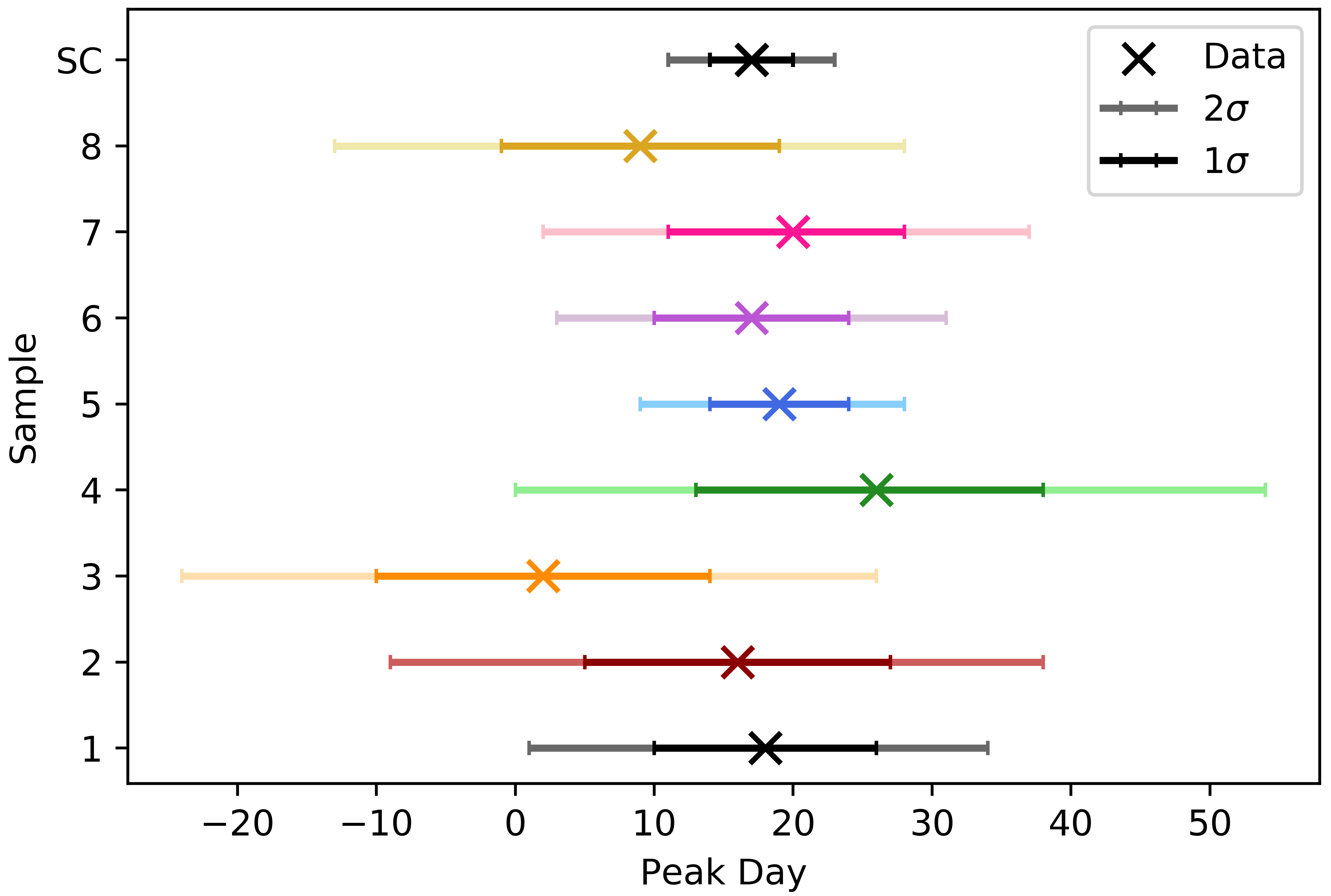}
\caption{The peak day of dark matter counts according to the annual modulation curves of the High Mass Model dark matter particle interacting with a Sodium Iodide detector. Evaluated for nuclear recoil energies with 1,2$\sigma$ errors.}
\label{fig:NaI_highmass_peakday_nuclear}
\end{figure}

Figure \ref{fig:am_NaI_highmass_2-6} demonstrates the annual modulation curve for a Sodium Iodide detector interacting with the High Mass Model dark matter particle. The result demonstrates an increase in modulation amplitude by a factor of over 4.5, compared to evaluating at an electron equivalent range of the same  value.

Figure \ref{fig:NaI_highmass_peakday_nuclear} demonstrates the expected peak day. The large uncertainties present make it difficult to constrain the peak day for this detector and dark matter candidate in the 2-6\keV energy region.

\bsp	
\label{lastpage}
\end{document}